\documentstyle[epsf,12pt]{article}

\font\elevenbf=cmbx10 scaled\magstep 1

\def\R3l{$R_3(l)$\ }
\def\R3h{$R_3(had)$\ }

\newbox\leftpage \newdimen\fullhsize \newdimen\hstitle \newdimen\hsbody
\catcode`\@=11 
\def\bigans{b }

\newcount\yearltd\yearltd=\year\advance\yearltd by -1900

\def\draftmode{\message{ DRAFTMODE }\def\draftdate{{\rm preliminary draft:
\number\month/\number\day/\number\yearltd\ \ \hourmin}}%
\headline={\hfil\draftdate}\writelabels\baselineskip=20pt plus 2pt minus 2pt
 {\count255=\time\divide\count255 by 60 \xdef\hourmin{\number\count255}
  \multiply\count255 by-60\advance\count255 by\time
  \xdef\hourmin{\hourmin:\ifnum\count255<10 0\fi\the\count255}}}
\def\nolabels{\def\wrlabeL##1{}\def\eqlabeL##1{}\def\reflabeL##1{}}
\def\writelabels{\def\wrlabeL##1{\leavevmode\vadjust{\rlap{\smash%
{\line{{\escapechar=` \hfill\rlap{\sevenrm\hskip.03in\string##1}}}}}}}%
\def\eqlabeL##1{{\escapechar-1\rlap{\sevenrm\hskip.05in\string##1}}}%
\def\reflabeL##1{\noexpand\llap{\noexpand\sevenrm\string\string\string##1}}}
\nolabels
\global\newcount\secno \global\secno=0
\global\newcount\meqno \global\meqno=1
\def\newsec#1{\global\advance\secno by1\message{(\the\secno. #1)}
\global\subsecno=0\eqnres@t\noindent{\bf\the\secno. #1}
\writetoca{{\secsym} {#1}}\par\nobreak\medskip\nobreak}
\def\eqnres@t{\xdef\secsym{\the\secno.}\global\meqno=1\bigbreak\bigskip}
\def\sequentialequations{\def\eqnres@t{\bigbreak}}\xdef\secsym{}
\global\newcount\subsecno \global\subsecno=0
\def\subsec#1{\global\advance\subsecno by1\message{(\secsym\the\subsecno. #1)}
\ifnum\lastpenalty>9000\else\bigbreak\fi
\noindent{\it\secsym\the\subsecno. #1}\writetoca{\string\quad
{\secsym\the\subsecno.} {#1}}\par\nobreak\medskip\nobreak}
\def\appendix#1#2{\global\meqno=1\global\subsecno=0\xdef\secsym{\hbox{#1.}}
\bigbreak\bigskip\noindent{\bf Appendix #1. #2}\message{(#1. #2)}
\writetoca{Appendix {#1.} {#2}}\par\nobreak\medskip\nobreak}
\def\eqnn#1{\xdef #1{(\secsym\the\meqno)}\writedef{#1\leftbracket#1}%
\global\advance\meqno by1\wrlabeL#1}
\def\eqna#1{\xdef #1##1{\hbox{$(\secsym\the\meqno##1)$}}
\writedef{#1\numbersign1\leftbracket#1{\numbersign1}}%
\global\advance\meqno by1\wrlabeL{#1$\{\}$}}
\def\eqn#1#2{\xdef #1{(\secsym\the\meqno)}\writedef{#1\leftbracket#1}%
\global\advance\meqno by1$$#2\eqno#1\eqlabeL#1$$}
\newskip\footskip\footskip14pt plus 1pt minus 1pt 
\setbox\strutbox=\hbox{\vrule height9.5pt depth4.5pt width0pt}
\global\newcount\ftno \global\ftno=0
\def\foot{\global\advance\ftno by1\footnote{$^{\the\ftno}$}}
\newwrite\ftfile
\def\footend{\def\foot{\global\advance\ftno by1\chardef\wfile=\ftfile
$^{\the\ftno}$\ifnum\ftno=1\immediate\openout\ftfile=foots.tmp\fi%
\immediate\write\ftfile{\noexpand\smallskip%
\noexpand\item{f\the\ftno:\ }\pctsign}\findarg}%
\def\footatend{\vfill\eject\immediate\closeout\ftfile{\parindent=20pt
\centerline{\bf Footnotes}\nobreak\bigskip\input foots.tmp }}}
\def\footatend{}
\global\newcount\refno \global\refno=1
\newwrite\rfile
\def\ref{[\the\refno]\nref}
\def\nref#1{\xdef#1{[\the\refno]}\writedef{#1\leftbracket#1}%
\ifnum\refno=1\immediate\openout\rfile=refs.tmp\fi
\global\advance\refno by1\chardef\wfile=\rfile\immediate
\write\rfile{\noexpand\item{#1\ }\reflabeL{#1\hskip.31in}\pctsign}\findarg}
\def\findarg#1#{\begingroup\obeylines\newlinechar=`\^^M\pass@rg}
{\obeylines\gdef\pass@rg#1{\writ@line\relax #1^^M\hbox{}^^M}%
\gdef\writ@line#1^^M{\expandafter\toks0\expandafter{\striprel@x #1}%
\edef\next{\the\toks0}\ifx\next\em@rk\let\next=\endgroup\else\ifx\next\empty%
\else\immediate\write\wfile{\the\toks0}\fi\let\next=\writ@line\fi\next\relax}}
\def\striprel@x#1{} \def\em@rk{\hbox{}}
\def\lref{\begingroup\obeylines\lr@f}
\def\lr@f#1#2{\gdef#1{\ref#1{#2}}\endgroup\unskip}

\def\addref#1{\immediate\write\rfile{\noexpand\item{}#1}} 

\def\startrefs#1{\immediate\openout\rfile=refs.tmp\refno=#1}
\def\xref{\expandafter\xr@f}\def\xr@f[#1]{#1}
\def\refs#1{\count255=1[\r@fs #1{\hbox{}}]}
\def\r@fs#1{\ifx\und@fined#1\message{reflabel \string#1 is undefined.}%
\nref#1{need to supply reference \string#1.}\fi%
\vphantom{\hphantom{#1}}\edef\next{#1}\ifx\next\em@rk\def\next{}%
\else\ifx\next#1\ifodd\count255\relax\xref#1\count255=0\fi%
\else#1\count255=1\fi\let\next=\r@fs\fi\next}
\newwrite\ffile\global\newcount\figno \global\figno=1
\def\fig{fig.~\the\figno\nfig}
\def\nfig#1{\xdef#1{fig.~\the\figno}%
\writedef{#1\leftbracket fig.\noexpand~\the\figno}%
\ifnum\figno=1\immediate\openout\ffile=figs.tmp\fi\chardef\wfile=\ffile%
\immediate\write\ffile{\noexpand\medskip\noexpand\item{Fig.\ \the\figno. }
\reflabeL{#1\hskip.55in}\pctsign}\global\advance\figno by1\findarg}
\def\xfig{\expandafter\xf@g}\def\xf@g fig.\penalty\@M\ {}
\def\figs#1{figs.~\f@gs #1{\hbox{}}}
\def\f@gs#1{\edef\next{#1}\ifx\next\em@rk\def\next{}\else
\ifx\next#1\xfig #1\else#1\fi\let\next=\f@gs\fi\next}
\newwrite\lfile
{\escapechar-1\xdef\pctsign{\string\%}\xdef\leftbracket{\string\{}
\xdef\rightbracket{\string\}}\xdef\numbersign{\string\#}}

\def\writestop{\def\writestoppt{\immediate\write\lfile{\string\pageno%
\the\pageno\string\startrefs\leftbracket\the\refno\rightbracket%
\string\def\string\secsym\leftbracket\secsym\rightbracket%
\string\secno\the\secno\string\meqno\the\meqno}\immediate\closeout\lfile}}
\def\writestoppt{}\def\writedef#1{}
\def\seclab#1{\xdef #1{\the\secno}\writedef{#1\leftbracket#1}\wrlabeL{#1=#1}}
\def\subseclab#1{\xdef #1{\secsym\the\subsecno}%
\writedef{#1\leftbracket#1}\wrlabeL{#1=#1}}
\newwrite\tfile \def\writetoca#1{}
\def\leaderfill{\leaders\hbox to 1em{\hss.\hss}\hfill}
\def\writetoc{\immediate\openout\tfile=toc.tmp
   \def\writetoca##1{{\edef\next{\write\tfile{\noindent ##1
   \string\leaderfill {\noexpand\number\pageno} \par}}\next}}}

\catcode`\@=12 

\def\inv{^{\raise.15ex\hbox{${\scriptscriptstyle -}$}\kern-.05em 1}}

\def\Dsl{\,\raise.15ex\hbox{/}\mkern-13.5mu D} 
\def\dsl{\raise.15ex\hbox{/}\kern-.57em\partial}

\def\lspace{\ifx\answ\bigans{}\else\qquad\fi}
\def\lbspace{\ifx\answ\bigans{}\else\hskip-.2in\fi} 
\def\darr#1{\raise1.5ex\hbox{$\leftrightarrow$}\mkern-16.5mu #1}
\def\ha{{1\over2}}
\def\roughly#1{\raise.3ex\hbox{$#1$\kern-.75em\lower1ex\hbox{$\sim$}}}

\begin{document}

\newcommand{\mc}{Monte Carlo }
\newcommand{\mcs}{Monte Carlos }
\newcommand{\brem}{brems\-strah\-lung }
\newcommand{\bq}{\begin{equation}}
\newcommand{\eq}{\end{equation}}
\newcommand{\bqa}{\begin{eqnarray}}
\newcommand{\eqa}{\end{eqnarray}}
\newcommand{\nn}{\nonumber \\}
\newcommand{\mpmm}{\mu^{+}\mu^{-}}
\newcommand{\tptm}{\tau^{+}\tau^{-}}
\newcommand{\into}{\rightarrow}
\newcommand{\sq}{^{2}}
\newcommand{\etal}{\it et al.\rm}
\newcommand{\ra}{\rightarrow}
\newcommand{\lnf}{{\ifmmode \Lambda^{(N_f)} \else $\Lambda^{(N_f)}$\fi}}
\newcommand{\ms}{{\ifmmode \overline{MS} \else $\overline{MS}$\fi}}
\newcommand{\dr}{{\ifmmode \overline{DR} \else $\overline{DR}$\fi}}
\newcommand{\lms}{{\ifmmode \Lambda^{(5)}_{\overline{MS}}
                 \else $\Lambda^{(5)}_{\overline{MS}}$\fi}}
\newcommand{\lam}{{\ifmmode \Lambda \else $\Lambda$\fi}}
\newcommand{\gev}{{\ifmmode {\rm GeV/c^2} \else ${\rm GeV/c^2}$\fi}}
\newcommand{\lp}{{\ifmmode L^+  \else $L^+$\fi}}
\newcommand{\lm}{{\ifmmode L^-  \else $L^-$\fi}}
\newcommand{\mlp}{{\ifmmode M(L^-)  \else $M(L^-)$\fi}}
\newcommand{\mlz}{{\ifmmode M(L^0)  \else $M(L^0)$\fi}}
\newcommand{\lz}{{\ifmmode L^0     \else $L^0$\fi}}
\newcommand{\ev}{{\ifmmode GeV/c^2       else $GeV/c^2$\fi}}
\newcommand{\tri}{{\ifmmode \triangleup  \else $\triangleup$\fi}}
\newcommand{\unl}{{\ifmmode U_{lL^0}  \else $U_{lL^0}$\fi}}
\newcommand{\gL}{{\ifmmode g_L  \else $g_{L}$\fi}}
\newcommand{\gR}{{\ifmmode g_R  \else $g_{R}$\fi}}
\newcommand{\gumu}{{\ifmmode \gamma^{\mu}  \else $\gamma^{\mu}$\fi}}
\newcommand{\gunu}{{\ifmmode \gamma^{\nu}  \else $\gamma^{\nu}$\fi}}
\newcommand{\gdmu}{{\ifmmode \gamma_{\mu}  \else $\gamma_{\mu}$\fi}}
\newcommand{\gdnu}{{\ifmmode \gamma_{\nu}  \else $\gamma_{\nu}$\fi}}
\newcommand{\stw}{{\ifmmode\sin^2\theta_W  \else $\sin^{2}\theta_{W}$
\fi}}
\newcommand{\sw}{{\ifmmode \sin^2\theta_W  \else $\sin^{2}\theta_{W}$
\fi}}
\newcommand{\swb}{{\ifmmode \sin^2\theta_{\overline{MS}}
                     \else $\sin^2\theta_{\overline{MS}}$
\fi}}
\newcommand{\cwb}{{\ifmmode \cos^2\theta_{\overline{MS}}
                     \else $\cos^2\theta_{\overline{MS}}$
\fi}}
\newcommand{\qq}{{\ifmmode q\overline{q} \else $q\overline{q}$\fi}}
\newcommand{\as}{{\ifmmode \alpha_s  \else $\alpha_s$\fi}}
\newcommand{\lR}{{\ifmmode l_R  \else $l_R$\fi}}
\newcommand{\lL}{{\ifmmode l_L  \else &l_L$\fi}}
\newcommand{\nt}{{\ifmmode \nu_{\tau} \else $\nu_{\tau}$\fi}}
\newcommand{\nuR}{{\ifmmode \nu_R  \else $\nu_R$\fi}}
\newcommand{\nuL}{{\ifmmode \nu_L  \else $\nu_L$\fi}}
\newcommand{\qR}{{\ifmmode g_R  \else $q_R$\fi}}
\newcommand{\qL}{{\ifmmode q_L  \else $q_L$\fi}}
\newcommand{\qRp}{{\ifmmode q_R'  \else $q_{R}$'\fi}}
\newcommand{\qLp}{{\ifmmode q_L'  \else $q_{L}$'\fi}}
\newcommand{\est}{{\ifmmode e^{\bf \ast}  \else $e^{\bf \ast}$\fi}}
\newcommand{\lst}{{\ifmmode l^{\bf \ast}  \else $l^{\bf \ast}$\fi}}
\newcommand{\must}{{\ifmmode \mu^{\bf \ast}  \else $\mu^{\bf \ast}$\fi}}
\newcommand{\taust}{{\ifmmode \tau^{\bf \ast}  \else $\tau^{\bf \ast}$
\fi}}
\newcommand{\pperp}{{\ifmmode p_t  \else $p_t$\fi}}
\newcommand{\et}{{\ifmmode E_t  \else $E_t$\fi}}
\newcommand{\xt}{{\ifmmode x_t  \else $x_t$\fi}}
\newcommand{\smumu}{{\ifmmode \sigma_{\mu\mu}  \else $\sigma_{\mu\mu}$
\fi}}
\newcommand{\eg}{{\ifmmode e\gamma  \else $e\gamma$\fi}}
\newcommand{\epem}{{\ifmmode e^+e^-  \else $e^+e^-$\fi}}
\newcommand{\lplm}{{\ifmmode L^+L^-  \else $L^+L^-$\fi}}
\newcommand{\pp}{{\ifmmode p\overline p  \else $p\overline p$\fi}}
\newcommand{\llz}{{\ifmmode L^0\overline{L}^0 \else
$L^0\overline{L}^0$\fi}}
\newcommand{\epemt}{{\ifmmode e^+e^- \to  \else $e^+e^- \to$\fi}}
\newcommand{\eb}{{\ifmmode E_{beam}  \else $E_{beam}$\fi}}
\newcommand{\ip}{{\ifmmode pb^{-1}  \else $pb^{-1}$\fi}}
\newcommand{\upm}{{\ifmmode ^{\pm}  \else $^{\pm}$\fi}}
\newcommand{\de}{{\ifmmode ^{\circ}  \else $^{\circ}$ \fi}}
\newcommand{\appr}{{\ifmmode \sim \else $\sim$ \fi}}
\newcommand{\corresp}{{\ifmmode \stackrel{\wedge}{=}
                      \else   $\stackrel{\wedge}{=}$ \fi}}
\newcommand{\sqrts}{{\ifmmode \sqrt{s} \else $\sqrt{s}$\fi}}
\newcommand{\zz}{{\ifmmode Z^0  \else $Z^0$\fi}}
\newcommand{\mz}{{\ifmmode M_{Z}  \else $M_{Z}$\fi}}
\newcommand{\mw}{{\ifmmode M_{W}  \else $M_{W}$\fi}}
\newcommand{\mh}{{\ifmmode M_{Higgs}  \else $M_{Higgs}$\fi}}
\newcommand{\gt}{{\ifmmode \Gamma_{tot} \else $\Gamma_{tot}$\fi}}
\newcommand{\msusy}{{\ifmmode M_{SUSY}  \else $M_{SUSY}$\fi}}
\newcommand{\taup} {{\ifmmode \tau_{proton} \else $\tau_{proton}$\fi}}
\newcommand{\agut}{{\ifmmode \alpha_{GUT}  \else $\alpha_{GUT}$\fi}}
\newcommand{\mgut}{{\ifmmode M_{GUT}  \else $M_{GUT}$\fi}}
\newcommand{\mze} {{\ifmmode m_0        \else $m_0$\fi}}
\newcommand{\mha}{{\ifmmode m_{1/2}    \else $m_{1/2}$\fi}}
\newcommand{\mb} {{\ifmmode m_{b}    \else $m_{b}$\fi}}
\newcommand{\mt} {{\ifmmode m_{t}    \else $m_{t}$\fi}}
\newcommand{\tb} {{\ifmmode \tan\beta  \else $\tan\beta$\fi}}
\newcommand {\rb}[1]{\raisebox{1.5ex}[-1.5ex]{#1}}

\hyphenation{multi-pli-ci-ties}
\hyphenation{cor-rections}


 \newlength{\dinwidth}
 \newlength{\dinmargin}
 \setlength{\dinwidth}{21.0cm}
 \textheight23.0cm \textwidth15.0cm
 \setlength{\dinmargin}{\dinwidth}
 \addtolength{\dinmargin}{-\textwidth}
 \setlength{\dinmargin}{0.5\dinmargin}
 \oddsidemargin -0.9in
 \addtolength{\oddsidemargin}{\dinmargin}
 \setlength{\evensidemargin}{\oddsidemargin}
 \setlength{\marginparwidth}{0.9\dinmargin}
 \marginparsep 8pt \marginparpush 5pt
 \topmargin -42pt
 \headheight 12pt
 \headsep 30pt \footheight 12pt \footskip 24pt
%
%
\newdimen\shrinkdim
\ifx\ans\bigans\shrinkdim=-2in\else\shrinkdim=-1.5in\fi
\global\newcount\tblno \global\tblno=1
%
%
\def\mytopinsert#1{
\break
\vbox{#1}\nobreak\bigskip}
%
\def\tbl{\ntbl}
\def\ntbl#1#2#3{\xdef#1{Table~\the\tblno}%
\bigskip\nopagebreak
\centerline{\vbox{#2}}
\medskip
\centerline{\hfill
\hbox {\bf Table \the\tblno:}
\vtop{\advance\hsize by \shrinkdim\baselineskip=14pt\noindent #3}
\hfill} \global\advance\tblno by1%
\bigskip
}%
%

\def\ai{\alpha_i}
\def\aii{\alpha_i^{-1}}
\def\rZ{{\rm Z}}
\def\rW{{\rm W}}
\def\rG{{\rm GUT}}
\def\rt{{\rm threshold}}
\def\rS{{\rm SUSY}}
\def\rH{{\rm Higgs}}
\def\rF{{\rm Fam}}
\def\MG{M_\rG}
\def\MS{M_\rS}
\def\MZ{M_\rZ}
\def\MW{M_\rW}
\def\Mt{M_\rt}
\def\MSbar{{\overline{MS}}}
\def\DRbar{{\overline{DR}}}
\medskip
\begin{flushright}
        IEKP-KA/93-13    \\
        hep-ph/9308238   \\
        August, 1993     \\
\end{flushright}

\vglue 2.0cm
\begin{center}{{\bf 
               Constraints on SUSY Masses in
               \vglue 3pt
Supersymmetric Grand Unified Theories           \\}
\vglue 1.0cm
{\bf    W.  de Boer\footnote{Bitnet: DEBOERW@CERNVM}
and R. Ehret\footnote{Bitnet: BD21@DKAUNI2}\\}
\baselineskip=13pt
{\it Inst.\ f\"ur Experimentelle Kernphysik, Univ.\ of Karlsruhe   \\}
\baselineskip=12pt
{\it Postfach 6980, D-76128 Karlsruhe 1, FRG                         \\}
\vglue 0.6cm
{\bf D.I. Kazakov            \\}
\baselineskip=13pt
{\it Joint Institute for Nuclear Research, Dubna   \\                }
\vglue 1.cm
{\bf    ABSTRACT}}
\end{center}
\vglue 0.3cm
{\rightskip=3pc
 \leftskip=3pc
 \rm\baselineskip=12pt
 \noindent

\begin{center}\parbox{13cm}{\small
Within the Minimal Supersymmetric Grand Unified Theory (MSGUT) masses
of the predicted supersymmetric particles are constrained by
the world averaged values of the electroweak and strong coupling
constants, the lower limits on the proton lifetime,
the lower limit on the
lifetime of the universe, which implies an upper limit on
the dark matter density,
the electroweak  symmetry breaking   originating  from radiative
corrections due to the heavy top quark, and the ratio of the  masses
of the b-quark and $\tau$-lepton.
A combined fit  shows that indeed
the MSGUT model can satisfy all these constraints simultaneously and
the corresponding values for
               all SUSY  masses are given within the minimal model,
taking into account the complete second order renormalisation
group equations for the couplings and the one-loop corrections
to the Higgs potential for the calculation of the \mz~mass and the
Higgs masses. These   one-loop corrections to \mz~ have been
 derived explicitly
 as function of the stop- and top masses and found to be
small for the best solution, but unnaturally large for the
90\% C.L. upper limits on the SUSY masses.
}
\vglue 0.8cm
(Contribution to the XVI International Symposium on \\
Lepton-Photon Interactions, \\
 Cornell, 10-15 August,
1993)
\end{center}
}
\lref\bek{
W. de Boer, R. Ehret and D.I. Kazakov, to be published}
\lref\lep{ 
        {\em Contributions from the LEP Coll. at the XXVI
International Conference on High Energy Physics}, Dallas, August 1992.
        }
\lref\WZ{
Yu.A. Gol'fand, E.P. Likhtman, JETP Lett. {\bf 13} (1971) 323; \\
D.V. Volkov, V.P. Akulow, Phys. Lett. {\bf 46b} (1971) 323; \\
J. Wess, B. Zumino, Nucl. Phys. {\bf B70} (1974) 39;\\
         For further references see the review papers:   \\
         H.-P. Nilles, Phys. Rep. {\bf 110} (1984) 1;\\
         H.E. Haber, G.L. Kane, Phys. Rep. {\bf 117} (1985) 75;\\
         A.B. Lahanas and D.V. Nanopoulos, Phys. Rep. {\bf 145} (1987) 1;
  \\     R. Barbieri, Riv Nuo. Cim. {\bf 11} (1988) 1.}

\lref\erz{J. Ellis, G. Ridolfi, F. Zwirner, Phys. Lett. {\bf B262}
(1991) 477\\
H.E. Haber R. Hempfling, Phys. Rev. Lett. {\bf 66} (1991) 83; \\
J. R. Espinosa, M. Quiros, Phys. Lett. {\bf B266} (1991) 389;
Z. Kunszt and F. Zwirner, Nucl. Phys. {\bf B 385} (1992) 3}
\lref\oamaldi{U. Amaldi, A. B\"ohm, L. S. Durkin, P. Langacker,
A. K. Mann, W. J. Marciano, A. Sirlin, H. H. Williams,
Phys. Rev. {\bf D36} (1987) 1385; \\
G. Costa, J. Ellis, G.L. Fogli, D.V. Nanopoulos, F. Zwirner, Nucl. Phys.
{\bf B297} (1988).}
\lref\pdb{Review of Particle Properties, Phys. Rev. {\bf D45} (1992). }
\lref\adf{U. Amaldi, W. de Boer, H. F\"urstenau, Phys. Lett. {\bf B260}
(1991) 447.}
\lref\adfI{U. Amaldi et al.,
                Phys. Lett. {\bf B281}
(1992) 374.}
\lref\rosI{
L. E. Ib\'a\~nez, G. G. Ross,
Nucl. Phys. {\bf B368} (1992) 3  and references therein}
\lref\rrb{G.G. Ross and R.G. Roberts, Nucl. Phys. {\bf B377} (1992) 571
           }
\lref\rob{G.G. Roberts and Roszkowski,  RAL-93-003
           }
\lref\nan{J.L. Lopez, D.V. Nanopoulos, and H. Pois, Phys. Rev.
{\bf D47} (1993) 2468.}
\lref\nanI{J.L. Lopez, D.V. Nanopoulos, and A. Zichichi,
CERN-TH.6903/93, and references therein.}
\lref\arn{Arnowitt and Nath, Phys. Rev. Lett. {\bf 69} (1992) 725;\\
Phys. Lett.  {\bf B287} (1992) 89;
Phys. Lett.  {\bf B289} (1992) 368; CTP-TAMU-39/92 (1992),
and references therein}
\lref\sac{A.D. Sakharov, ZhETF Pis'ma {\bf 5} (1967) 32
           }
\lref\pic{
              E. Braaten S. Narison, A.Pich,
              CERN-TH 6070/91 (1991);\\
                A. Pich, {\em Invited talk at Recontres de Moriond},
Les Arc, May 1992, CERN-TH 6489/92 (1992).
           }
\lref\rev{%
G. Altarelli, talk at  the Workshop ``QCD: 20 years later'',
Aachen, Germany, 1992, CERN Preprint CERN-TH-6623-92; \\
S. Bethke, Nucl. Phys. {\bf A546} (1992) 247;  \\
S. Bethke and J.E. Pilcher,
Ann. Rev. Nucl. Part. Sci 42 (1992) 251; \\
W. de Boer,  Proc. of the  18th Slac Summer Institute on Particle
Physics, July 1990, p. 431; \\
T. Hansl-Kozanecka, to be published in the Proc. of the 19th Slac
Summer Institute on Particle Physics, July, 1991,
Curie Univ. of Paris VI, preprint LPNHE-92-03; \\
T. Hebbeker, Phys. Rep. 217 (1992) 69.
}
\lref\resum{
                 S. Bethke and  S. Catani,
{\em Invited talks at Recontres de Moriond}, Les Arc, May 1992,
CERN-TH.8484/92.
           }
\lref\dis{
CCFR Coll., J.R. Mishra et al., Nevis Preprints 1459,1460,1461 (1992)\\
CDHSW Coll., P. Berge et al., Z. Phys. {\bf C49} (1991) 187\\
CHARM Coll., F. Bergsma et al., Phys. Lett. {\bf 153B} (1985) 111\\
A.D. Martin, R.G. Roberts, W.J. Stirlin,
Phys. Lett. {\bf B266} (1991) 273; \\
M. Virchaux, A. Milsztajn, Phys. Lett. {\bf B274} (1992) 221.
           }
\lref\uaone{
 A. Geiser, {\em Invited talk at Recontres de Moriond},
Les Arc, May 1992, Aachen preprint PITHA 91/19 (1992);\\
M. Lindgren et al., Univ. of California Preprint UCR/UA1/91-01 (1991)
}
\lref\uatwo{
UA2 Collab., J. Alitti et al., Phys. Lett. {\bf 263} (1991) 563.
           }
\lref\svi{
               W. de Boer and T. Ku\ss maul,
Karlsruhe preprint
{\bf IEKP-KA/92-11} (1992). 
           }
\lref\kob{
                       M. Kobel, {\em Invited talk at Recontres de
Moriond},
Les Arc, May 1992, CERN-TH.6489/92 (1992).
           }
\lref\dal{
       R. Tanaka and G. Rolandi, {\em Invited talks at the XXVI
International Conference on High Energy Physics}, Dallas, August 1992.
           }
\lref\cosm{The early Universe by G. B\"orner, Springer Verlag (1991)\\
          The early Universe by
E.W. Kolb and M.S. Turner, Addison-Wesley(1990)\\
A. Guth and P. Steinhardt in
The new Physics, edited by P. Davis, Cambridge University Press
(1989), p34.
           }
\lref\dfs{G. Degrassi, S. Fanchiotti, A. Sirlin, Nucl. Phys.
{\bf B351} (1991) 49.}
%

\lref\fg{P. H. Frampton, S. L. Glashow, Phys. Lett. {\bf 131B} (1983) 340;
E: {\bf 135B} (1984) 515.}
\lref\einjon{M. B. Einhorn, D. R. T. Jones, Nucl. Phys. {\bf B196} (1982)
475; \\
             M. E. Machacek, M. T. Vaughn, Nucl. Phys. {\bf B222}
(1983) 83.}
\lref\lanluo{P. Langacker, M. Luo, Phys. Rev. {\bf D44} (1991) 817.}
\lref\lanI{P. Langacker, N. Polonski, Univ. of Pennsylvania Preprint
UPR-0556-T, (1993)}
\lref\lanII{P. Langacker,  Univ. of Pennsylvania Preprint
UPR-0539-T, (1993)}    
\lref\pok{M. Olechowski and S. Pokorski, Max-Planck-Institute
Preprint MPI-PH-92-118 (1992)}
\lref\car{M. Carena, S. Pokorski, C.E.M. Wagner, Max-Planck-Institute
Preprint MPI-PH-93-10 (1993), and private communication}
\lref\aslep{
                          J. Lefran\c cois,
 Int. Eur. Conf. on High Energy Physics,    Marseille, July, 1993}
\lref\mar{ G. Altarelli,
 Int. Eur. Conf. on High Energy Physics,    Marseille, July, 1993}
\lref\aslepI{   T. Hebbeker, Phys. Rep. {\bf 217} (1992) 69;\\
S. Bethke, Plenary talk at the XXVI International Conf. on
High Energy Physics, Dallas (USA), August 1992, Heidelberg Preprint
HD/PY 92-13\\
G. Altarelli, Proceedings of the  Conf. "QCD-20 Years later",
Aachen, Germany  June 1992, CERN-TH.6623/92   }
\lref\weinberg{S. Weinberg, Phys. Lett. {\bf 91B} (1980) 51.}
\lref\akt{I. Antoniadis, C. Kounnas, K. Tamvakis, Phys. Lett. {\bf 119B}
(1982) 377.}
\lref\ekn{J. Ellis, S. Kelley, D. V. Nanopoulos, Phys. Lett. {\bf B260}
(1991) 131.}
\lref\eknII{J. Ellis, S. Kelley, D. V. Nanopoulos,
Nucl. Phys. {\bf B373} (1992) 55}
\lref\baha{
R. Barbieri, L.J. Hall, Phys. Rev. Lett. {\bf (68} (1992) 752
 }
\lref\gggqw{H. Georgi, S. L. Glashow, Phys. Rev. Lett. {\bf 32} (1974) 438\\
H. Georgi, H. R. Quinn, S. Weinberg, Phys. Rev. Lett. {\bf 33} (1974) 451.}
\lref\MSSMref{
P. Fayet, Phys. Lett. {\bf B64} (1976) 159; ibid. {\bf B60}
(1977) 489; \\
S. Dimopoulos, H. Georgi, Nucl. Phys. {\bf B193} (1981) 150; \\
L. E. Ib\'a\~nez, G. G. Ross, Phys. Lett. {\bf B105} (1981) 435; \\
S. Dimopoulos, S. Raby, F. Wilczek, Phys. Rev. {\bf D24} (1981) 1681.}

\lref\spart{A. B. Lahanas, D. V. Nanopoulos, Phys. Rep. {\bf 145} (1987) 1\\
G. G. Ross,                    {\it Proceedings of Joint International
Lepton-Photon Symposium and Europhysics Conference on High Energy Physics},
Geneva, 25 July -- 1 August, 1991.}
\lref\bbdm{W. A. Bardeen, A. Buras, D. Duke, T. Muta, Phys. Rev. {\bf D18}
(1978) 3998.}
\lref\acpz{F. Anselmo, L. Cifarelli, A. Petermann, A. Zichichi,
Il Nuovo Cimento {\bf 104A} (1991) 1817.
}
\lref\aczt{F. Anselmo, L. Cifarelli,  A. Zichichi,
   Il Nuovo Cimento {\bf 105} (1992) 1335; ibid. {\bf 105} (1992) 1357
and references therein.}
\lref\aczpt{F. Anselmo, L. Cifarelli, A. Peterman, A. Zichichi,
   Il Nuovo Cimento {\bf 105} (1992) 1179}
\lref\bmas{
L. E. Ib\'a\~nez and C. L\'opez, Phys. Lett. {\bf 126B} (1983) 54;
Nucl. Phys. {\bf B233} (1984) 511}
\lref\mtmax{B. Pendleton and G.G. Ross, Phys. Lett.
{\bf B98} (1981) 291; \\ C.T. Hill, Phys. Rev. {bf D24} (1981) 691}
\lref\gas{J. Gasser and H. Leutwyler, Phys. Rep. {\bf 87C} (1982) 77;\\
S. Narison, Phys. Lett. {\bf B216} (1989) 191}
\lref\yana{
H. Murayama, T. Yanagida, Preprint Tohoku University, TU-370 (1991);\\
T.G. Rizzo, Phys. Rev. {\bf D45} (1992) 3903
         }

 \clearpage
{}.
\newpage
\newsec{Introduction}
Grand Unified Theories (GUT) hold the promise of "explaining"
 the difference between the electromagnetic, weak and strong nuclear
 forces: their different strenghts are simply due to radiative
 corrections.
Furthermore, they are candidates to explain several unrelated
observations about our universe, e.g. they almost automatically
lead to baryon number violation, thus providing a possible explanation
for the matter-antimatter asymmetry  in our universe\sac\
and the                       spontaneous symmetry breaking of the
unified force into the known forces at a sufficient high  energy
                         can cause the inflationary scenario of the
universe, thus providing an explanation for
           the origin of matter and the homogeneity
of the universe on a large scale.
The interested reader is refered to
 recent text books on this exciting subject \cosm\
for more details and original references.

         The  $SU(5)$ group, which is the
smallest group encompassing the $SU(3)$ and $SU(2)\bigotimes U(1)$
groups of the strong- and electroweak interactions,
can be ruled out as a viable GUT,   since it predicts too
rapid proton decay. An upper limit on
the proton lifetime can be        estimated
from the unification scale and any GUT is required to have
the unification scale above $10^{15}$ GeV in order to be compatible
with the proton life time limits. This is not the case in the SU(5)
model and these limits severely constrain other   GUT models.

Indeed, the coupling constants, as measured precisely
at LEP, do not unify either within the $SU(5) $ model, i.e.
they do not become equal at a single energy
if extrapolated to high energies, but
 within the supersymmetric
extension of the $SU(5)$ model (MSSM)\MSSMref,
unification is obtained\ekn\adf\adfI\lanluo\eknII.
Supersymmetry\WZ\ presupposes a symmetry between fermions and bosons,
thus introducing spin 0 partners of the quarks and leptons
-- called squarks and sleptons -- and spin 1/2 partners of
the gauge bosons and  Higgs particles -- called gauginos and Higgsinos.
Since these predicted particles have not been observed sofar,
these supersymmetric (SUSY) particles
must be heavier than the known particles, implying that
supersymmetry must be broken.
However, from the unification condition a first estimate of
 the SUSY breaking scale could be made: it was found to
 be of the order of 1000 GeV, or more precisely $10^{3\pm1}$  GeV\adf.
The uncertainty in this scale is mainly caused
by the uncertainty in the strong coupling constant.

Clearly, the whole SUSY mass spectrum cannot be described
by a single parameter. In the MSSM one needs at least 5
parameters. So many parameters cannot be derived from the
unification condition alone.
 However,
further constraints can be considered:
\begin{itemize}
\item
\mz~predicted from electroweak  symmetry breaking\rrb\rosI.
\item
b-quark mass predicted from the unification of
Yukawa couplings\bmas\rosI\lanI.
\item
Constraints from the lower limit on the proton lifetime\arn\nan\lanII.
\item
Constraints from the lower limit on the lifetime of the universe\rob.
\item
The maximum value of the top mass is restricted by the couplings\rosI.
\item
          {Experimental lower limits on SUSY masses\lep.}
\end{itemize}
All these constraints have been considered separately or
partly\ekn\adf\adfI\lanluo\eknII\nan\pok\nanI.
However,
 considering only one constraint at a time allows one to obtain   only
one relation between parameters.  Trying to find complete solutions
 requires additional assumptions, like naturalness,
 no-scale models, fixed ratios for gaugino- and scalar masses
 or a fixed ratio  for the Higgs mixing parameter  and the
 scalar mass, assumptions from supergravity, or
 combinations of these assumptions.

The different assumptions lead sometimes to apparently
conflicting results. For example, the
dark matter constraint  requires the mass of the scalar
particles at \mgut~to be below 500 GeV\rob, while the analysis
of the proton life time constraint requires this mass
to be above 600 GeV\arn.

It is the purpose of this paper to study all constraints
simultaneously without any of the assumptions mentioned above
in order to see if a solution within the {\it minimal}
supersymmetric model (MSSM) exists at all, which is non-trivial
as should be clear from the contradictions mentioned before.

 It turns out that a solution exists indeed and the
 corresponding constraints
on the SUSY mass spectrum, the strong coupling constant and the
top mass are given, if all constraints from the experimental data
mentioned above
are imposed simultaneously (for the first time as fas as we know).

\newsec{Experimental constraints  }
\subsec{Unification of the couplings}

In the unified $\rm SU(2)\bigotimes U(1)$ theory,
the following well known
tree-level relations hold between the couplings
and the gauge boson masses
%
\eqn\SW{\matrix{
e&=&\sqrt{4\pi\alpha}=g\sin\theta_W=g'\cos \theta_W\cr
\MW&=&\ha v g \cr
\MZ&=&\ha v\sqrt{g^{\prime2}+g^2}\cr}        }
%
%
%
from which it follows that
\eqn\sinthW{\sin^2\theta_W={e^2\over g^2}={g^{\prime2}\over g^{\prime2}+g^2}
=1-{\MW^2\over\MZ^2}}
Here $g$ and $g'$ are the couplings of the groups SU(2) and U(1)
respectively, $\alpha$ is the fine structure constant and $v$ is the
 vacuum expectation value of the
Higgs field.  If the model contains Higgs representations other than
doublets, the theory has an additional degree of freedom, usually
parametrized by the $\rho$ parameter.

In the SM based on the group $\rm SU(3)\times SU(2)\times U(1)$ we use the
usual definitions of the couplings
\eqn\SMcoup{\matrix{
\alpha_1&=&(5/3)g^{\prime2}/(4\pi)&=&5\alpha/(3\cos^2\theta_W)\cr
\alpha_2&=&\hfill g^2/(4\pi)&=&\alpha/\sin^2\theta_W\hfill\cr
\alpha_3&=&\hfill g_s^2/(4\pi)\cr}}
where $g_s$ is the SU(3) coupling.  The factor of $5/3$ in the
definition of $\alpha_1$ has been included for the proper normalization at
the unification point\gggqw.  The couplings, when defined as
effective values including loop corrections in the gauge boson propagators,
become energy dependent (``running'').  A running coupling requires
the specification of a renormalization prescription, for which one usually
uses the modified minimal subtraction ($\MSbar$) scheme\bbdm.

In this scheme the world averaged values of the couplings at the
Z$^0$ energy are
\eqn\worave{\matrix{
\alpha^{-1}(\MZ)&=&127.9\pm0.2\cr
\sin^2\theta_\MSbar&=&0.2326\pm0.0006\cr
\alpha_3&=&0.123\pm0.008\cr}               }
The value of $\alpha^{-1}$ is given in Ref. \dfs\ and the
    value of $\sin^2\theta_\MSbar$ has been presented at Marseille\mar.
The $\alpha_3$ value
corresponds to value measured at LEP form the ratio  of the
hadronic- and leptonic cross sections\aslep, which
agrees well with the world average\rev.
               This value  has the smallest theoretical
uncertainty from the higher order corrections,
we prefer  to take this value alone.

For SUSY models, the dimensional reduction $\DRbar$ scheme is a more
appropriate renormalization scheme\akt.  This scheme also has the advantage
that all thresholds can be treated by simple step approximations.  Thus
unification occurs in the $\DRbar$ scheme if all three $\aii(\mu)$ meet
exactly at one point.
                     This crossing point then gives the mass of the heavy
gauge bosons.  The $\MSbar$ and $\DRbar$ couplings differ by a small offset
\eqn\MSDR{{1\over\alpha_i^\DRbar}={1\over\alpha_i^\MSbar}-{C_i\over\strut12\pi}
}
where the $C_i$ are the quadratic Casimir coefficients of the group ($C_i=N$
for SU($N$) and 0 for U(1) so $\alpha_1$ stays the same).  Throughout the
following, we use the $\DRbar$ scheme for the MSSM.

The energy dependence of the couplings is completely determined by
the particle content and their couplings inside the loop diagrams of the
gauge bosons as expressed by the renormalization group (RG) equations.
The RG equations can be rewritten as
\eqn\RGinv{{d\over d\ln \mu}\aii(\mu)={-1\over2\pi}\left(
b_i+\sum_{j=1}^3{b_{ij}\over4\pi}\alpha_j(\mu)+O(\alpha_j^2)\right).}
                In first order, i.e. all $b_{ij}=0$,
the equations for the three $\aii$ are
independent with a linear solution in the $\aii$---$\log\mu$ plane.
When the second order contributions are taken
into account, the equations become coupled and the running of each $\aii$
depends on the values of the other two couplings.  However, the
second order effects are small because of the additional factor
$\alpha_j/(4\pi)\le0.01$.  Higher orders are presumably even smaller by
additional powers of $\alpha_j/(4\pi)$.
We solve \RGinv\ by numerical integration.

\subsec{\mz~ constraint  from electroweak symmetry breaking}
In the MSSM at least two
Higgs doublets have   to be introduced:

$$ H_1(1,2,-\frac{1}{2})= \left(\begin{array}{c}H^0_1 \\
H^-_1\end{array}\right), \ \ \
 H_2(1,2,\frac{1}{2})= \left(\begin{array}{c}H^+_2 \\
H^0_2\end{array}\right),$$

At the tree level the interactions of the Higgs fields can be
parametrised by an effective potential of the form:
$$V(H_1,H_2)=m^2_1|H_1|^2+m^2_2|H_2|^2-m^2_3(H_1H_2+h.c.)+
\frac{g^2+g^{'2}}{8}(|H_1|^2-|H_2|^2)^2,$$

where $m_1$, $m_2$ and $m_3$ are mass parameters, for which
we assume at
 the GUT scale the following boundary conditions:
 $m_1^2=m^2_2=\mu^2_0+m_0^2, \ m^2_3=B\mu_0m_0$.

Radiative corrections
from the heavy top and stop quarks  can drive one of the Higgs masses
negative, thus causing  spontaneous
            symmetry breaking in the electroweak sector.
            In this case the Higgs potential does not have
            its minimum for all fields equal zero, but the
the minimum is obtained for non-zero vacuum expectation  values
of the fields:
$$<H_1>\equiv v_1=\frac{v}{\sqrt{2}}\cos\beta , \ \
<H_2>\equiv v_2=\frac{v}{\sqrt{2}}\sin\beta,                $$
$$ v^2=\frac{v_1^2+v_2^2}{2}, \ \  \tan\beta \equiv \frac{v_1}{v_2}$$

 The scale, where symmetry breaking occurs depends on the
 starting values of the mass parameters at the GUT scale,
the top mass and the evolution of the couplings and masses.
This gives strong constraints between the known \zz~mass and the
SUSY mass parameters, as demonstrated e.g. in Ref.\rrb.

Minimization of the tree level potential yields:
$$v^2=\frac{\displaystyle 8(m^2_1-m^2_2\tan^2
\beta )}{\displaystyle (g^2+g^{'2})(\tan^2\beta -1)},\ \ \
\sin2\beta =\frac{\displaystyle 2m^2_3}{\displaystyle m^2_1+m^2_2}$$
$$ M^2_W=\frac{g^2}{4}v^2, \ \ M^2_Z=\frac{g^2+g^{'2}}{4}v^2,$$
After including  the one-loop corrections to the potential\erz,
the \zz~mass becomes dependent on the top- and stop quark masses too.
In this case we derived the following expression:
\begin{eqnarray*}
 M^2_Z&=&2\frac{\displaystyle m^2_1-m^2_2 \tan^2\beta -
 \Delta^2_Z}{\tan^2\beta -1}, \\
 \Delta^2_Z&=&\frac{3g^2}{32\pi^2}\frac{m^2_t}{M^2_W}\left[
 f(\tilde{m}^2_{t1})+f(\tilde{m}^2_{t2})+2m^2_t +(A^2_t-\mu^2\cot^2\beta )
 \frac{f(\tilde{m}^2_{t1})-f(\tilde{m}^2_{t2})}{\tilde{m}^2_{t1}-
 \tilde{m}^2_{t2}}\right]\\
\end{eqnarray*}

where

$$ f(m^2)=m^2(\ln\frac{m^2}{m^2_t}-1), \ \
h(m^2)=\frac{m^2}{m^2-\tilde{m}^2_{q}}\ln \frac{m^2}{\tilde{m}^2_{q}} $$
$$ d(m^2_1,m^2_2)=2-\frac{m^2_1+m^2_2}{m^2_1-m^2_2}\ln
\frac{m^2_1}{m^2_2}$$
$\tilde{m}^2_q$ is the mass of a light squark and
$\tilde{m}^2_{1,2}$ the stop quark masses.
Note that the corrections $\Delta_Z$ are zero if the top- and stop
quark masses are identical, i.e. if supersymmetry would be exact.
They grow with the difference $\tilde{m}^2_t-\mt^2$, so these
corrections become unnaturally large for large values of the
stop masses, as will be discussed later.

The  masses of the physical  Higgs particles
after spontaneous symmetry breaking become after inclusion of
the one-loop corrections\erz:
\begin{eqnarray*}
m^2_A &= &\frac{\Delta^2_A}{\sin2\beta },\\
\Delta^2_A&=&2m^2_3-\frac{3g^2}{16\pi^2}\frac{m^2_tA_t\mu}{\sin^2\beta
M^2_W}\frac{f(\tilde{m}^2_{t1})-f(\tilde{m}^2_{t2})}{\tilde{m}^2_{t1}-
\tilde{m}^2_{t2}}\\
m^2_{H^{\pm}}&=&m^2_A+M^2_W+\Delta^2_H, \\
\Delta^2_H&=&-\frac{3g^2}{32\pi^2}\frac{m^2_t\mu^2}{\sin^4\beta
M^2_W}\frac{h(\tilde{m}^2_{t1})-h(\tilde{m}^2_{t2})}{\tilde{m}^2_{t1}-
\tilde{m}^2_{t2}}\\
m^2_{h,H}&= &
\frac{1}{2}\left[m^2_A+M^2_Z +\Delta_{11}+\Delta_{22} \right.\\ &&\left. \pm
\sqrt{\begin{array}{ll}
(m^2_A+M_Z^2+\Delta_{11}+\Delta_{22})^2 & -4m^2_AM_Z^2\cos^22\beta
-4(\Delta_{11}\Delta_{22}-\Delta_{12}^2)\\ -4(\cos^2\beta
M^2_Z+\sin^2\beta M^2_A)\Delta_{22} & -4(\sin^2\beta M^2_Z+\cos^2\beta
M^2_A)\Delta_{11}\\ -4\sin2\beta (M^2_Z+M^2_A)\Delta_{12}& \end{array} }\right]
 \\
\Delta_{11}&=&\frac{3g^2}{16\pi^2}\frac{m^4_t}{\sin^2\beta M^2_W}
\left[\frac{\mu(A_t+\mu\cot\beta )}{\tilde{m}^2_{t1}-\tilde{m}^2_{t2}}
\right]^2d(\tilde{m}^2_{t1},\tilde{m}^2_{t2}), \\
\Delta_{22}&=&\frac{3g^2}{16\pi^2}\frac{m^4_t}{\sin^2\beta M^2_W}\left[
\ln (\frac{\tilde{m}^2_{t1}\tilde{m}^2_{t2}}{m^4_t})+
\frac{2A_t(A_t+\mu\cot\beta )}{\tilde{m}^2_{t1}-\tilde{m}^2_{t2}} \ln
(\frac{\tilde{m}^2_{t1}}{\tilde{m}^2_{t2}}) \right.\\
 & & \left. \left[
+\frac{A_t(A_t+\mu\cot\beta
)}{\tilde{m}^2_{t1}-\tilde{m}^2_{t2}}\right]^2
d(\tilde{m}^2_{t1},\tilde{m}^2_{t2})\right], \\
\Delta_{12}&=&\frac{3g^2}{16\pi^2}\frac{m^4_t}{\sin^2\beta M^2_W}
\frac{\mu(A_t+\mu\cot\beta )}{\tilde{m}^2_{t1}-\tilde{m}^2_{t2}} \left[
\ln (\frac{\tilde{m}^2_{t1}}{\tilde{m}^2_{t2}}) +
\frac{A_t(A_t+\mu\cot\beta
)}{\tilde{m}^2_{t1}-\tilde{m}^2_{t2}}
d(\tilde{m}^2_{t1},\tilde{m}^2_{t2})\right], \\
& & \\
A(t)&=&\frac{\displaystyle A_0}{\displaystyle 1+6Y_0F(t)}
+\frac{m_{1/2}}{m_0}\left(H_2-
\frac{\displaystyle 6Y_0H_3}{\displaystyle 1+6Y_0F(t)}\right) \\
\end{eqnarray*}
 Here $H_2, H_3, E$ and $F$ are functions of the couplings as defined in \rosI.
\subsec{Evolution of the masses}
In the soft breaking term of the Lagrangian \mze~ and \mha~ are the
universal masses of the gauginos and scalar particles at the
GUT scale, respectively and $\mu$ determines the masses
of the particles in the Higgs sector.
At lower energies the masses of the SUSY particles
start to differ from these  universal masses
due to the radiative corrections. E.g. the coloured particles
get  contributions proportional to $\as^2$ from gluon loops,
while the non-coloured ones get contributions depending on
the electroweak coupling constants only.
The evolution of the masses is given by the renormalisation group
equations\rosI.
We have used analytical solutions\bek\ of these equations
including the top Yukawa coupling, the Higgs mass parameters,
mass mixing between the top quarks, mixing between neutralinos,
mixing between the charginos and one loop radiative corrections
to the Higgs potential.
\subsec{b-quark mass constraint}
Unification of the Yukawa couplings for a given generation at the
GUT scale predicts relations for quark and lepton masses
within a given family.
This does not work for the light quarks, but the
ratio of b-quark and $\tau$-lepton masses can be correctly
predicted by the radiative corrections to the masses\rosI.
\subsec{Proton lifetime constraints}
GUT's predict proton decay and the present lower limits
on the proton lifetime yield quite strong  constraints
on the GUT scale and the SUSY parameters.
As mentioned at the beginning, the direct decay $p\rightarrow e^+\pi^0$
via s-channel exchange requires
the GUT scale to be above $10^{15}$ GeV. This is not fulfilled
in the SM, but always fulfilled in the MSSM. Therefore we do not
consider this constraint.
However, the decay via box diagrams with winos and Higgsinos
predict much shorter lifetimes, especially in the preferred mode
 $p\rightarrow \overline{\nu} K^+$.
 From the present experimental lower limit of  $10^{32}$ yr for
this decay mode Arnowitt and Nath\arn\  deduce an upper limit
on the parameter B:  $$B<293\pm 42 (M_{H_3}/3\mgut)\ GeV^{-1}$$
Here $M_{H_3}$ is the Higgsino mass, which is expected to be
of the order of $\mgut$. To obtain a conservative upper limit
on $B$, we allow  $M_{H_3}$ to become an order of magnitude heavier
than $\mgut$, so we require $$B< 977\pm 140\ GeV^{-1}.$$
 The   uncertainties from the unknown heavy Higgs mass  are large
compared with the contributions from the first and third generation,
which contribute through the mixing in the CKM matrix.
Therefore   we only consider the second order generation
contribution, which can be written as\arn\ :
$$B=-2(\alpha_2/(\alpha_3  \sin(2\beta))(m_{\tilde{g}}/
          m^2_{\tilde{q}}) ~10^6$$
One observes that the upper limit on $B$ favours small
 gluino masses  $m_{\tilde{g}}$, large squark masses
         $ m_{\tilde{q}} $, and small values of \tb.
         To fulfill   this constraint requires
                    $$\tb < 5$$ for the whole parameter space.
                              Arnowitt and Nath note that
         requiring the gluino mass to be below 500 GeV implies
         the mass of the scalar particles (\mze) at the GUT scale
         to be above 600 GeV.  We will not impose this
         requirement on the gluino mass.
Furthermore, they require $M_{H_3} <3~\mgut$,  so they obtained
tighter limits on \tb, since we allow $M_{H_3}<10~\mgut$.
\subsec{t-quark mass constraints}
For large Yukawa couplings the masses become dependent only
on the couplings (pole term in the RGE).
This yields a maximum value of the top mass, which has to be higher
than the experimental mass, thus       constraining the
SUSY parameters.
More precisely, the top mass can be expressed as:
$$\mt^2=(4\pi)^2\ Y_t(t)\ v^2\ \sin^2(\beta)$$
where the running of the Yukawa coupling
as function of $t=log(\frac{M_X^2}{Q^2})$ is given by\rosI:
$$Y_t(t)=\frac{\displaystyle Y_0E(t)}{\displaystyle 1+6Y_0F(t)}$$
One observes that for large Yukawa couplings $Y_0$ at the GUT scale,
        $Y_t$ becomes independent of $Y_0$ and the maximum value
of $\mt$ becomes:
$$\mt^2=\frac{(4\pi)^2\ E(t)}{6F(t)}\ v^2\ \sin^2(\beta),$$
where $E$ and $F$ are functions of the couplings only\rosI.
Clearly the experimental values of $\mt$ have to be below
this upper bound, which
 is most  easily fulfilled for larger values of $\tb$.
However, the proton life time limits require $\tb < 5$, as discussed
before. In this case  the upper limit on $\mt$ implies a  constraint
on the ratio  of $E/F$,  i.e. on the
starting point  at the GUT scale and the intermediate SUSY thresholds.
\subsec{Constraints from the lifetime of the universe}
The lightest supersymmetric particle (LSP) is supposedly stable
and would be  an ideal candidate for dark  matter.
So the MSSM predicts dark matter. However, from the long lifetime
of the universe one knows that the density of the universe
cannot be higher than the critical density, which implies
most of the LSP's have annihilated into photons.
This can only happen fast enough if   the
squarks and sleptons are sufficiently light, thus
posing a strong upper limit on some of the SUSY mass parameters.
Requiring the density of the universe to be below the critical density
translates into  an upper bound of about 500 GeV on  \mze~
for a large range of \mha\rob, i.e. $$\mze < 500~GeV.$$

\subsec{Experimetal lower limits on SUSY masses}
SUSY particles have not been found sofar and from the searches
at LEP one knows that the lower limit on the charged particles is
about half the \zz~ mass (45 GeV) and the Higgs mass has to be above
60 GeV\lep.
This requires also minimal values for the SUSY mass parameters.
\newsec{Constrained Fits}
\subsec{Fit strategy}
As mentioned before, given the 5  parameters in the MSSM and   $\agut$
and $\mgut$, all other SUSY masses, the b-quark mass, \mz,
 and the proton lifetime can be calculated.
In addition, the complete evolution of the
 couplings including all thresholds
 can be performed.
Furthermore, the dark matter constraint requires
\mze\ to     be below 500 GeV.

Therefore we have adopted  the following strategy:
we varied \mze\ between 0 and 500 GeV and fitted the remaining 5
parameters: $\agut,\ \mgut,\ \mha,\ \mu,\ $ and $\tb$.
The trilinear coupling $A_0$ in the Higgs potential  at $\mgut$
was kept mostly at 0, but the large radiative
corrections to it were taken into account for the top quarks.
Varying $A_0$ between $+\mze$ and $-\mze$ did not change the
results significantly, so it was kept zero for the results quoted
hereafter.

The remaining parameters were fitted with MINUIT by minimizing the
following $\chi^2$ function:

\begin{eqnarray*} \chi^2&=&
       {\sum_{i=1}^3\frac{(\aii(\mz)-\alpha^{-1}_{MSSM_i}(\mz))^2}{
\sigma_i^2}}      \\
 & &+\frac{(\mz-91.18)^2}{\sigma_Z^2}     \\
 & &+\frac{(\mb-4.25)^2}{\sigma_b^2}     \\
 & &+\frac{(\mt-\mt^{max})^2}{\sigma_t^2}  {(for~\mt>\mt^{max})}\\
 & &+{\frac{(B  - 997)^2}{\sigma_B^2}} {(for ~B > 997)}  \\
 & &+{\frac{(D(m1m2m3))^2}{\sigma_D^2}} {(for~ D > 0)}     \\
 & &+{\frac{(\tilde{M}-\tilde{M}_{exp})^2}{\sigma_{\tilde{M}}^2}}
 {(for~\tilde{M} < \tilde{M}_{exp})}\\
\end{eqnarray*}
The first term is the contribution of the difference between the
calculated and measured coupling  constants at \mz~and  the following
three terms the contributions from the \mz-mass, \mb--mass, and
\mt--mass constraints. The last three terms give the contributions
from the proton lifetime, the requirement of electroweak symmetry
breaking, i.e. $D=m_1^2 m_2^2 -m_3^4 < 0$,
and experimental lower limits on the SUSY masses.
The following errors were attributed:
$\sigma_i$ equal the experimental errors in the coupling constants,
as defined before, $\sigma_b$=0.3 GeV,
 $\sigma_B$=0.14 GeV, and all the other errors were set to 10 GeV.
 The value of these errors turned out not to be critical at all,
 since the corresponding terms in the numerator
 were usually zero
 in case of a good fit and even for the 90\% C.L.  values
 these constraints could be fulfilled and the $\chi^2$ was determined
 by  the other terms, for which we know the errors.

For unification in the $\DRbar$ scheme,
 all three couplings $\aii(\mu)$ must
cross at a single unification point in
the $\aii$---$\mu$ plane given by $\MG$
and $\alpha_\rG^{-1}$ (the inverse of the unified coupling).  Thus in
these models
we can fit the couplings at $\mz$ by extrapolation from a single
point at $\MG$ back to $\mz$ for each of the $\alpha_i$'s and taking
into account all thresholds. Between the highest SUSY threshold
and $\mz$ only the first order coefficients
in the RGE  are known
for the individual  thresholds, at least  as far as we know.
However, the second order coefficients
must be between the values of the  MSSM including all particles
and the SM. So we varied these  second order coefficients in this range
for the small region between $\mz$ and the highest SUSY mass.
The difference was found to be negligible. Of course, for
the large extrapolation between the highest SUSY mass and $\mgut$
the complete second order RGE              was used.

\subsec{Results     }
The upper part of
Fig. 1  shows the evolution of the coupling constants
in the MSSM for two cases: one for the minimum value of the $\chi^2$
function as defined above (solid lines) and one corresponding to
the 90\% C.L. upper limit of the thresholds    of the light
SUSY particles (dashed lines).
The  light         thresholds  are indicated in the lower
function as the change in the first order coefficient in the
$\beta$ function, which corresponds to the first order change
in the slopes of the curves   at the top.
One observes that the change  in $\alpha_3$   occurs in a rather
narrow energy regime, corresponding to the threshold of squarks
and gluinos, while for the other coupling constants the sleptons,
higgsinos and winos contribute in addition, thus causing
a somewhat more smeared threshold region.


The parameters corresponding to these fits are tabulated
in Table 1.    The initial choices of $\mze$
as well as the fitted paramters are shown at the bottom.
As mentioned before, varying $A_0$ between $+\mze$ and $-\mze$
does not influence the results very much, so it was kept at 0,
but the large radiative corrections to it were taken into account.
The remaining parameters do not depend strongly on the   initial
choices of  \mze~ and \mt~, as shown in  Fig. 2.
The value of $\mze$  was varied between 0 and 500 GeV
and $\mt$
between 140 and 170  GeV. Larger values of $\mze$ are not allowed
by the dark matter constraint and the range of $\mt$ is
the preferred range from
 the b-quark mass, as
              shown in Fig. 3.
              The difference between the two bent lines
originates from the difference in \agut~and \mgut,
as found from the fits for different input values of \mt.
The horizontal band corresponds to the mass of the b-quark
after QCD corrections: $\mb=4.25\pm 0.1$ GeV\gas.
Although the error of 0.1 GeV is the quoted one, we used
the more conservative error estimate of 0.3 GeV, since
the value of the mass depends on the value of the strong coupling
constant at this mass, which is not too well known.
Note that the allowed range of \mt~ is in excellent
agreement with LEP results too\lep.

In order to obtain 90\% C.L. upper limits on the parameters, we varied
them until the $\chi^2$ value increased to 1.64.
Of course, this depends on the errors, so we used the rather
conservative  errors  defined above.
 The upper limit on $\tb$ was found to be about 5; here
the main constraint is
coming from the proton lifetime.

Since \mze~ is rather small, the solutions are not strongly dependent
on it, but they are mainly determined by $\mu$ and $\mha$. Since also
 $\tb$ is constrained     to a rather narrow range, we
are left effectively with two parameters, \mha~ and $\mu$.
However,   they
                          were found to be strongly
 correlated, as shown if Fig. 4.
 In the minimum the $\chi^2$ value is zero, but one notices
 a long valley,   where the $\chi^2$ is only slowly increasing.
 One can easily understand such a behaviour from Fig. 1:
 as long as the breakpoints in all curves move up simultaneously,
 unification can be obtained at  a higher value of $\agut$,
 as is obvious from a comparison of the dashed and solid lines.
 If one tries to combine the solid lines with the dashed lines
 into a single unification point, this clearly does not work.
 Consequently the breakpoints in all curves have to be close
 together in energy.
 However,  $\alpha_3$ is independent of  $\mu$  in contrast to
  the breakpoints in $\alpha_1$ and $\alpha_2$. This  implies
  that both $\mu$ and $\mha$ have to increase simultaneously
  in order to keep the breakpoints in all three curves
  close together.
  This strong correlation has been neglected in previous
  analysis, where $\mu$ and $\mha$ were choosen independently\rrb\aczpt.
 This clearly cannot work
  in a more detailed analysis, as observed too in Ref. \car.

 The steep walls in Fig. 4 originate from  the experimental lower
 limits on the SUSY masses and the value of \mz~ from
 radiative symmetry breaking.

Fig. 5 shows the evolution of the masses for the minimum value
of $\chi^2$ and $\mze$=500 GeV.
{}From Table 1 one observes that some SUSY masses can go up
to several TeV, if one considers the 90\% C.L.
Such large values spoil the cancellation
of the quadratic divergencies. If supersymmetry would be exact,
 i.e. as long as the masses and couplings of the particles and
 their superpartners are the same,
the contributions of fermions and bosons in the loops would
 exactly cancel each other, thus eliminating in first order
 all   divergencies.
 This can be seen explicitly in the corrections to $\mz$:
 $\Delta_Z$ is exactly zero if the masses of stop-- and top quarks
 are identical.
 For the  SUSY masses at the minimum value of $\chi^2$ the
corrections to $\mz$
are indeed very small, as shown in Fig. 6 on the
left hand side, but for the solutions corresponding to
the 90\% C.L. upper limit the corrections to $\mz$ are about
50 times $\mz$ itself (see Fig. 6 right hand side).
In addition, the value of the
determinant$^{1/4}$=$(m_1^2 m_2^2-m_3^4)^{1/4}$ in the Higgs potential
times its sign is shown as a dotted line. It clearly becomes negative,
which implies the potential takes the shape of the mexican hat.
The evolution of the masses in the Higgs potential is shown
in the     lower parts of Fig. 6.

If we require that only solutions are allowed, for which the
corrections to $\mz$ are not large compared with $\mz$ itself,
we have to limit the mass of the heaviest stop quark  to
about 1 TeV.
In this case the 90\% C.L. upper limits if the individual
SUSY particles are given in the right hand column if Table 1.
The correction to $\mz$ = 5 times \mz~ in this case.

To determine the lower  limits is somewhat more cumbersome,
since not all particles reach their minimum for the same set
of parameters.
We used the following strategy: each parameter was scanned for the
lowest possible value, while keeping the other parameters    free.
A zero value for the scalar mass
$\mze $ could  not be excluded, which corresponds to the so-called
{\it no-scale} model and the corresponding results have been
given in the left column of Table 1.
The parameters
$\mu$ and $\mha$  cannot    be choosen independently and their
minimal values were obtained for $\mze=400$ GeV, as shown in the
second column of Table 1. Lower  values would cause a wino mass
below the experimental lower limit of about 45 GeV.
The top mass for these minima was kept at 170 GeV, since smaller
values would give a too high value of \mb.

One observes from Table 1 the well known effect\erz
that the Higgs particle, which gets a negative mass squared before
spontaneous symmetry breaking (SSB),
  gets only a small mass after SSB.
The mass  of this particle,
   called $h$ in Table 1,
is a rather strong function of \mt,
as shown in Fig. 7.
For each value  of \mt, the  parameters   $\mha$,  $\mu$, and $\tb$
were determined by minimising the $\chi^2$
for the indicated  value of \mze.
One observes that the mass of the lightest Higgs particle varies
between 60 and 120 GeV. These values correspond to the minimal
value of $\chi^2$, but even if the 90\% C.L. limits are taken,
the mass  increases only to 125 GeV.
\newsec{Summary}


   The MSSM model has many predictions, which can be compared
with experiment, even in the energy range where the predicted
SUSY particles are out of reach. Among these predictions:

\begin{itemize}
\item
\mz.
\item
\mb.
\item
Proton decay.
\item
Dark Matter.
\item
Upper limit on \mt.
\end{itemize}
It is surprising, that the {\it minimal} supersymmetric model
can fulfil  all experimental constraints for these predictions.
As far as we know, supersymmetric   models are the only ones, which
are  consistent
 with all these observations simultaneously.
Other models can yield unification too\adfI, but they do not exhibit the
elegant symmetry properties of supersymmetry, they offer no
explanation for dark matter and no explanation for the electroweak
symmetry breaking. Furthermore the quadratic divergencies do not cancel.

{}From the above constraints we find at the 90\% C.L.
(see Table 1):
\begin{eqnarray*}
    0~ <& ~\mze~ &<~ 500~ \mbox{GeV}           \\
   70~ <& ~\mha~ &<~ 3410~ (475)~\mbox{GeV}    \\
  500~ <& ~\mu~  &<~ 3150~ (1000)~\mbox{GeV}    \\
    1~ <& ~\tb~  &<~ 5                         \\
  138~ <& ~\mt~  &<~ 186~\mbox{GeV}            \\
0.108~ <& ~\as~  &<~0.132                      \\
\end{eqnarray*}

    The upper limit on $\mze$ originates from the dark matter
    constraint, the upper limit on $\tb$ from the lower
    limit on the proton lifetime.
    The fact that $\tb$ is so much smaller than the ratio
    of top-- and b-quark mass implies that the Yukawa coupling
    of the b-quark is negligibly small, so one does not have
    to consider its contributions in the renormalisation
    group equations.

    Good fits are only obtained for $\as$ between 0.108 and 0.132,
    if the error on $\as$ is taken to be 0.008.
    The bottom mass constraint  together with the given couplings
    require   the top mass to be
    between 138 and 186 GeV.

The values in brackets indicate the 90\% C.L., if one requires
that the exact one-loop corrections to \mz~are not large compared
with \mz~itself, which requires the heaviest stop quark
to be below 1000 GeV. In this case the correction to \mz~ is
about 5 \mz and
the corresponding constraints on the other
SUSY masses are (see Table 1 for details):
\begin{eqnarray*}
 25~<&  \tilde{\gamma}(\chi^0_1)      &<~ 200 ~\mbox{GeV}  \\
 45~<&  \tilde{Z}(\chi^0_2) ,
   \tilde{W}(\chi^{\pm}_1)            &<~  385 ~\mbox{GeV} \\
 190~<& \tilde{g}                     &<~ 1110  ~\mbox{GeV} \\
 430~<& \tilde{q}                     &<~ 1075  ~\mbox{GeV} \\
 200~<& \tilde{t}_1                   &<~  730  ~\mbox{GeV} \\
 390~<& \tilde{t}_2                   &<~ 1000  ~\mbox{GeV} \\
 235~<& \tilde{e}_L                   &<~  520  ~\mbox{GeV} \\
 130~<& \tilde{e}_R                   &<~  440  ~\mbox{GeV} \\
 335~<& \tilde{H}                     &<~  785  ~\mbox{GeV} \\
 525~<&       {H}                     &<~  870  ~\mbox{GeV} \\
  60~<&       {h}                     &<~  125  ~\mbox{GeV} \\
\end{eqnarray*}

The values of $\mha$ and $\mu$ are positively correlated, i.e.
 a  large   (small) value of $\mha$ corresponds to a
    large   (small) value of $\mu$, as is apparent from Fig. 4.
This strong correlation was usually not taken into account
 in previous analysis,
in which $\mha$ and $\mu$ were         restricted by ad-hoc
assumptions\rrb\aczpt.

The detailed mass spectra  have
been given in Table 1.
The lightest Higgs particle  is certainly within reach of
experiments at present or
future accelerators. Its observation in the predicted mass range
of 60 to 125 GeV   would
be a strong case   in support of
    this minimal version of the supersymmetric grand unified theory.

\vglue 0.1cm
{\elevenbf\noindent Acknowledgments}
\vglue 0.2cm
We thank
Ugo Amaldi,    Hermann F\"urstenau, Stavros Katsanevas, Sergey Kovalenko,
R.G. Roberts  and F. Zwirner
for their interest in this work and helpful discussions.
\bigskip\\

\renewcommand{\arraystretch}{1.30}
\renewcommand{\rb}[1]{\raisebox{1.75ex}[-1.75ex]{#1}}

\begin{table}[htb]
\begin{center}
\begin{tabular}{|c|r|r||r||r|r|}  \hline
                  & \multicolumn{5}{|c|}{masses in $[$GeV$]$ } \\
\cline{2-6}
   \rb{particles}
&      \multicolumn{2}{|c|}{ \makebox[3.8cm]{Lower limit }}&
      \makebox[1.9cm]{{\bf best fit}} &
       \multicolumn{2}{|c|}{ \makebox[3.8cm]{Upper limit 90\% C.L. }} \\
\hline\hline
 fine tuning & \makebox[1.9cm]{ no } &\makebox[1.9cm]{ no } &
                \makebox[1.9cm]{ {\bf no }}& %
                \makebox[1.9cm]{ no }      & %
                \makebox[1.9cm]{ $m_{\tilde{t}} < 1$~TeV}\\
\hline\hline
  $\tilde{\gamma}(\chi^0_1)$     & 134 &23 &{\bf  46}& 1501  &  199   \\
\hline
  $\tilde{Z}(\chi^0_2)$          & 254&46&{\bf  88}  & 2850  &  383   \\
\hline
  $\tilde{W}(\chi^{\pm}_1)$      & 254&44&{\bf  87}  & 2900  &  385   \\
\hline
  $\tilde{g}$                    & 799 &193&{\bf 330}& 6826  & 1109   \\
\hline  \hline
  $\tilde{e}_L$                  & 236 &406&{\bf 412}& 2331  &  521   \\
\hline
  $\tilde{e}_R$                  & 131&402&{\bf 404} & 1384  &  441   \\
\hline
  $\tilde{\nu}_L$                & 228&399&{\bf406}  & 2331  &  516   \\
\hline  \hline
  $\tilde{q}_L$                  & 725&434&{\bf 497} & 6035  & 1075   \\
\hline
  $\tilde{q}_R$                  & 697&431&{\bf 490} & 5700  & 1036   \\
\hline
  $\tilde{t}_1$                  & 498&201&{\bf 226} & 4740  &  729   \\
\hline
  $\tilde{t}_2$                  & 748&391&{\bf 472} & 5600  & 1000   \\
\hline        \hline
  $\tilde{H}_1(\chi^0_3)$        & 550&321&{\bf 417}& 3311  &  771   \\
\hline
  $\tilde{H}_2(\chi^0_4)$        & 572&337&{\bf 436}& 3323  &  784   \\
\hline
  $\tilde{H}^{\pm}(\chi^{\pm}_2)$& 569&337&{\bf 433}& 3324  &  783   \\
\hline   \hline
  $       h $                    &  91&101&{\bf  96}&  115  &  127   \\
\hline
  $       H $                    & 624&525&{\bf 636}& 2897  &  871   \\
\hline
  $       A $                    & 619&523&{\bf 633}& 2897  &  870   \\
\hline
  $       H ^{\pm}$              & 625&529&{\bf 638}& 2897  &  873   \\
\hline\hline
 \multicolumn{6}{|c|}{ SUSY parameters } \\
\hline
 \mze          &  0 &400&{\bf 400} &  500 & 400               \\
\hline
 \mha          &329 &70 &{\bf 172} & 3413 & 475               \\
\hline
 $\mu$         &550 &507&{\bf 576} & 3150 & 1009              \\
\hline
 $\tan\beta$   &2.0 &3.7&{\bf 2.2} &  2.9 & 3.5               \\
\hline
 \mt           &147 &186&{\bf 172} &  138 & 175               \\
\hline
 1/\agut       &24.7&24.2&{\bf 24.3} & 26.4 & 25.1               \\
\hline
 \mgut         &$1.3\;10^{16}$&$2.0\;10^{16}$& $
{\bf 1.9\;10^{16}}$ & $0.7\;10^{16}$ & $1.2\;10^{16}$ \\
\hline
 \end{tabular}
\end{center}
 \caption{\label{t1} Values of SUSY masses and parameters.
 For the lower  limits of the SUSY masses one
 should take the lowest value from
 the first two columns, except for  the value of the lightest Higgs $h$.
 In that case the lowest value is given by Fig. 7.        }
\end{table}
%
%
\clearpage
\centerline{\bf References     }
 \begin{description}
  \setlength{\leftmargin}{3.7cm}
  \setlength{\listparindent}{0cm}
\item {[1]\ }%
A.D. Sakharov, ZhETF Pis'ma {\bf 5} (1967) 32
\item {[2]\ }%
The early Universe by G. B\"orner, Springer Verlag (1991)\\
The early Universe by
E.W. Kolb and M.S. Turner, Addison-Wesley(1990)\\
A. Guth and P. Steinhardt in
The new Physics, edited by P. Davis, Cambridge University Press
(1989), p34.
\item {[3]\ }%
P. Fayet, Phys. Lett. {\bf B64} (1976) 159; ibid. {\bf B60}
(1977) 489; \\
S. Dimopoulos, H. Georgi, Nucl. Phys. {\bf B193} (1981) 150; \\
L. E. Ib\'a\~nez, G. G. Ross, Phys. Lett. {\bf B105} (1981) 435; \\
S. Dimopoulos, S. Raby, F. Wilczek, Phys. Rev. {\bf D24} (1981) 1681.
\item {[4]\ }%
J. Ellis, S. Kelley, D. V. Nanopoulos, Phys. Lett. {\bf B260}
(1991) 131.
\item {[5]\ }%
U. Amaldi, W. de Boer, H. F\"urstenau, Phys. Lett. {\bf B260}
(1991) 447.
\item {[6]\ }%
U. Amaldi et al.,
Phys. Lett. {\bf B281}
(1992) 374.
\item {[7]\ }%
P. Langacker, M. Luo, Phys. Rev. {\bf D44} (1991) 817.
\item {[8]\ }%
J. Ellis, S. Kelley, D. V. Nanopoulos,
Nucl. Phys. {\bf B373} (1992) 55
\item {[9]\ }%
Yu.A. Gol'fand, E.P. Likhtman, JETP Lett. {\bf 13} (1971) 323; \\
D.V. Volkov, V.P. Akulow, Phys. Lett. {\bf 46b} (1971) 323; \\
J. Wess, B. Zumino, Nucl. Phys. {\bf B70} (1974) 39;\\
For further references see the review papers: \\
H.-P. Nilles, Phys. Rep. {\bf 110} (1984) 1;\\
H.E. Haber, G.L. Kane, Phys. Rep. {\bf 117} (1985) 75;\\
A.B. Lahanas and D.V. Nanopoulos, Phys. Rep. {\bf 145} (1987) 1;
\\ R. Barbieri, Riv Nuo. Cim. {\bf 11} (1988) 1.
\item {[10]\ }%
G.G. Ross and R.G. Roberts, Nucl. Phys. {\bf B377} (1992) 571
\item {[11]\ }%
L. E. Ib\'a\~nez, G. G. Ross,
Nucl. Phys. {\bf B368} (1992) 3 and references therein
\item {[12]\ }%
L. E. Ib\'a\~nez and C. L\'opez, Phys. Lett. {\bf 126B} (1983) 54;
Nucl. Phys. {\bf B233} (1984) 511
\item {[13]\ }%
P. Langacker, N. Polonski, Univ. of Pennsylvania Preprint
UPR-0556-T, (1993)
\item {[14]\ }%
Arnowitt and Nath, Phys. Rev. Lett. {\bf 69} (1992) 725;\\
Phys. Lett. {\bf B287} (1992) 89;
Phys. Lett. {\bf B289} (1992) 368; CTP-TAMU-39/92 (1992),
and references therein
\item {[15]\ }%
J.L. Lopez, D.V. Nanopoulos, and H. Pois, Phys. Rev.
{\bf D47} (1993) 2468.
\item {[16]\ }%
G.G. Roberts and Roszkowski, RAL-93-003
\item {[17]\ }%
 {\em Contributions from the LEP Coll. at the XXVI
International Conference on High Energy Physics}, Dallas, August 1992.
\item {[18]\ }%
H. Georgi, S. L. Glashow, Phys. Rev. Lett. {\bf 32} (1974) 438\\
H. Georgi, H. R. Quinn, S. Weinberg, Phys. Rev. Lett. {\bf 33} (1974) 451.
\item {[19]\ }%
W. A. Bardeen, A. Buras, D. Duke, T. Muta, Phys. Rev. {\bf D18}
(1978) 3998.
\item {[20]\ }%
G. Degrassi, S. Fanchiotti, A. Sirlin, Nucl. Phys.
{\bf B351} (1991) 49.
\item {[21]\ }%
R. Tanaka and G. Rolandi, {\em Invited talks at the XXVI
International Conference on High Energy Physics}, Dallas, August 1992.
\item {[22]\ }%
T. Hebbeker, Phys. Rep. {\bf 217} (1992) 69;\\
S. Bethke, Plenary talk at the XXVI International Conf. on
High Energy Physics, Dallas (USA), August 1992, Heidelberg Preprint
HD/PY 92-13\\
G. Altarelli, Plenary talk at the Conf. "QCD-20 Years later",
Aachen, Germany June 1992, CERN-TH.6623/92
\item {[23]\ }%
I. Antoniadis, C. Kounnas, K. Tamvakis, Phys. Lett. {\bf 119B}
(1982) 377.
\item {[24]\ }%
J. Ellis, G. Ridolfi, F. Zwirner, Phys. Lett. {\bf B262}
(1991) 477\\
H.E. Haber R. Hempfling, Phys. Rev. Lett. {\bf 66} (1991) 83; \\
J. R. Espinosa, M. Quiros, Phys. Lett. {\bf B266} (1991) 389;
Z. Kunszt and F. Zwirner, Nucl. Phys. {\bf B 385} (1992) 3
\item {[25]\ }%
W. de Boer, R. Ehret and D.I. Kazakov, to be published
\item {[26]\ }%
J. Gasser and H. Leutwyler, Phys. Rep. {\bf 87C} (1982) 77;\\
S. Narison, Phys. Lett. {\bf B216} (1989) 191
\item {[27]\ }%
F. Anselmo, L. Cifarelli, A. Peterman, A. Zichichi,
Il Nuovo Cimento {\bf 105} (1992) 1179
\item {[28]\ }%
M. Carena, S. Pokoroski, C.E.M. Wagner, Max-Planck-Institute
Preprint MPI-PH-93-10 (1993), and private communication

\end{description}
%
%
%
\begin{figure}
 \begin{center}
  \leavevmode
  \epsfxsize=14cm
  \epsfysize=18cm
  \epsffile{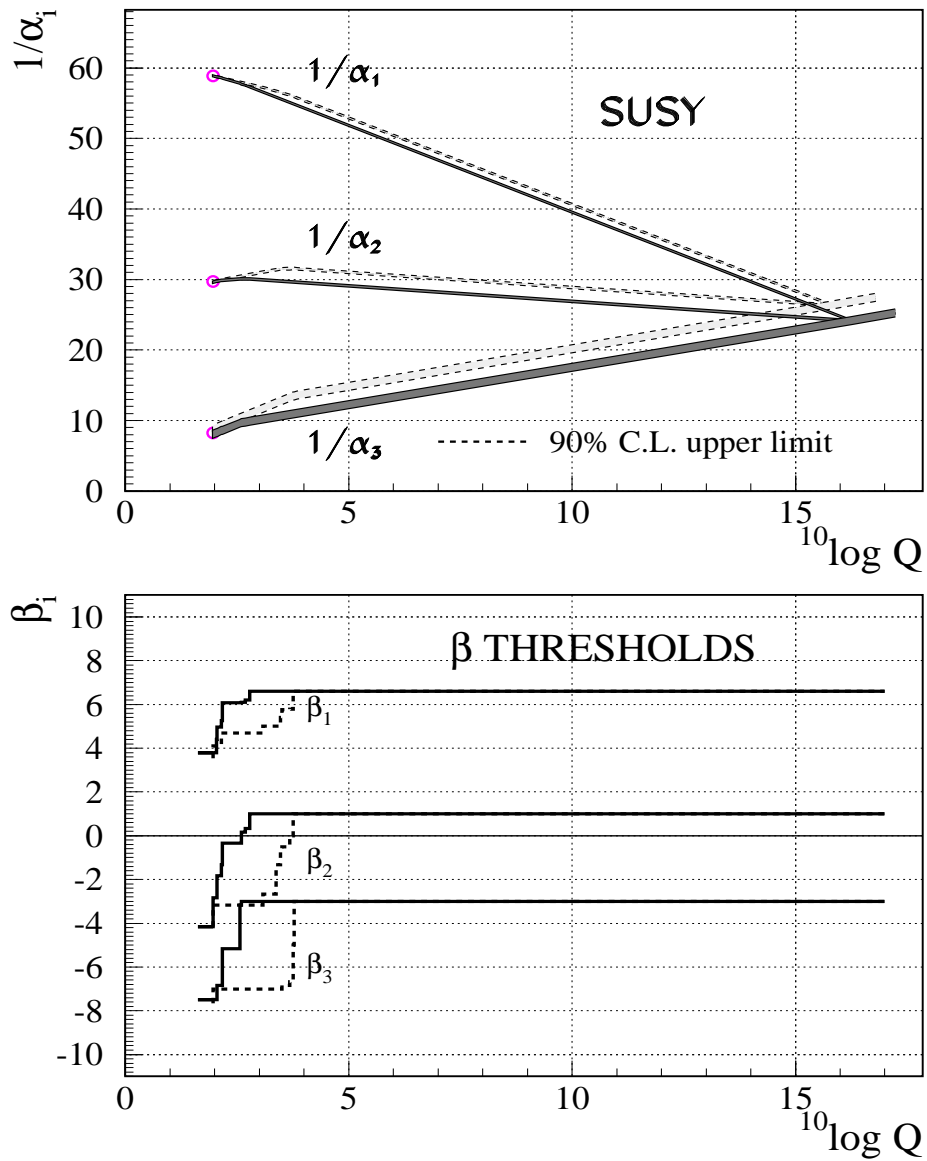}
 \end{center}
 \caption{\label{f1}
 Evolution of the inverse of the three couplings in the
 MSSM.                   The line  above $\MG $ follows the prediction
 from the supersymmetric SU(5) model.
The SUSY thresholds have been indicated in the lower part  of the curve:
they are treated as step functions in the
     first order $\beta$ coefficients in the renormalisation group
equations, which correspond to a change in slope in the evolution
of the couplings in the top figure.
The dashed lines correspond to the 90\% C.L. upper limit for the
SUSY thresholds.
}
\end{figure}
\clearpage

\begin{figure}
 \begin{center}
  \leavevmode
  \epsfxsize=14cm
  \epsfysize=18cm
  \epsffile{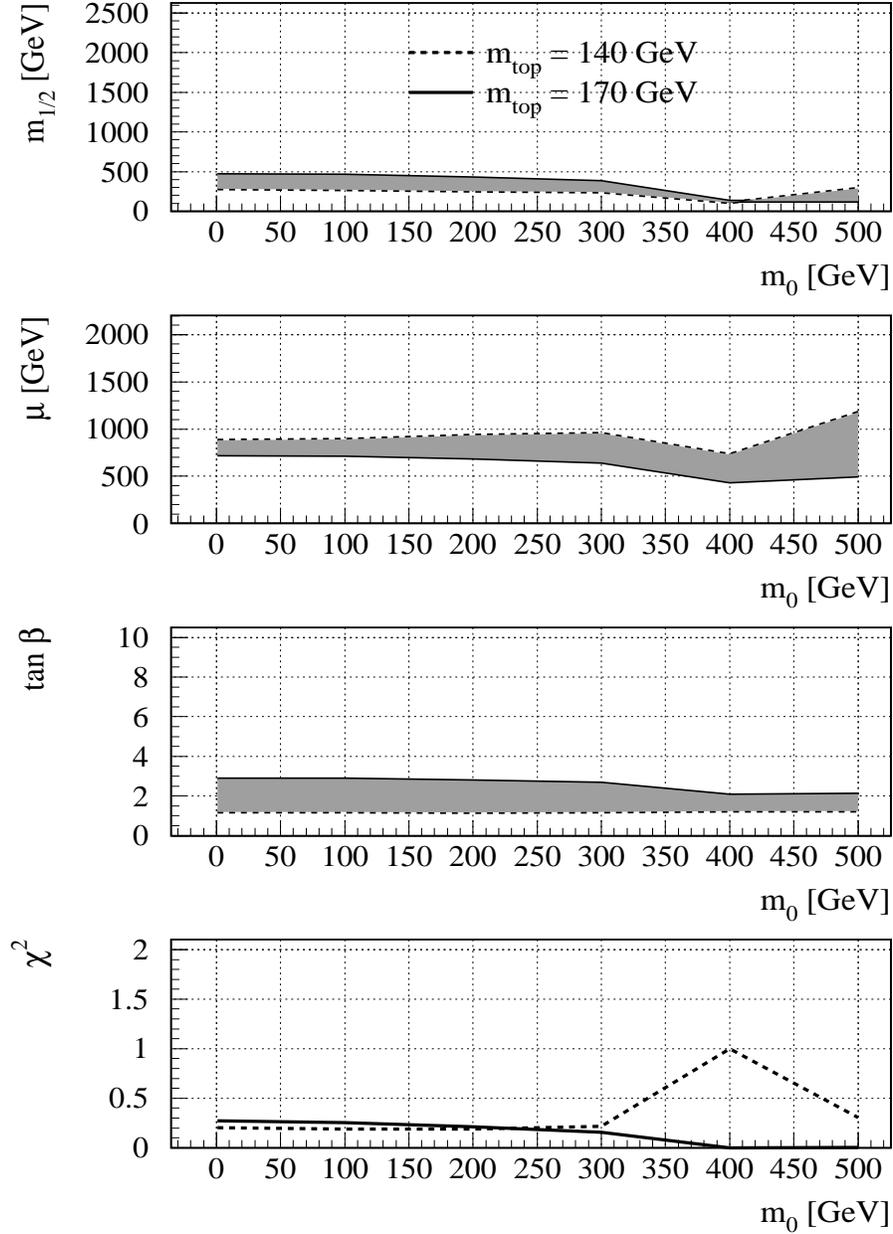}
 \end{center}

  \caption{\label{f2}
   The fitted MSSM parameters  as function of the choice of $\mze$:
   above $\mha$, middle $\mu$ and below
   $\tb$.
   The shaded  area indicates  the variation, if $\mt$ is varied
   between 140 and 170 GeV. }
\end{figure}
\clearpage

\begin{figure}
 \begin{center}
  \leavevmode
  \epsfxsize=14cm
  \epsfysize=18cm
  \epsffile{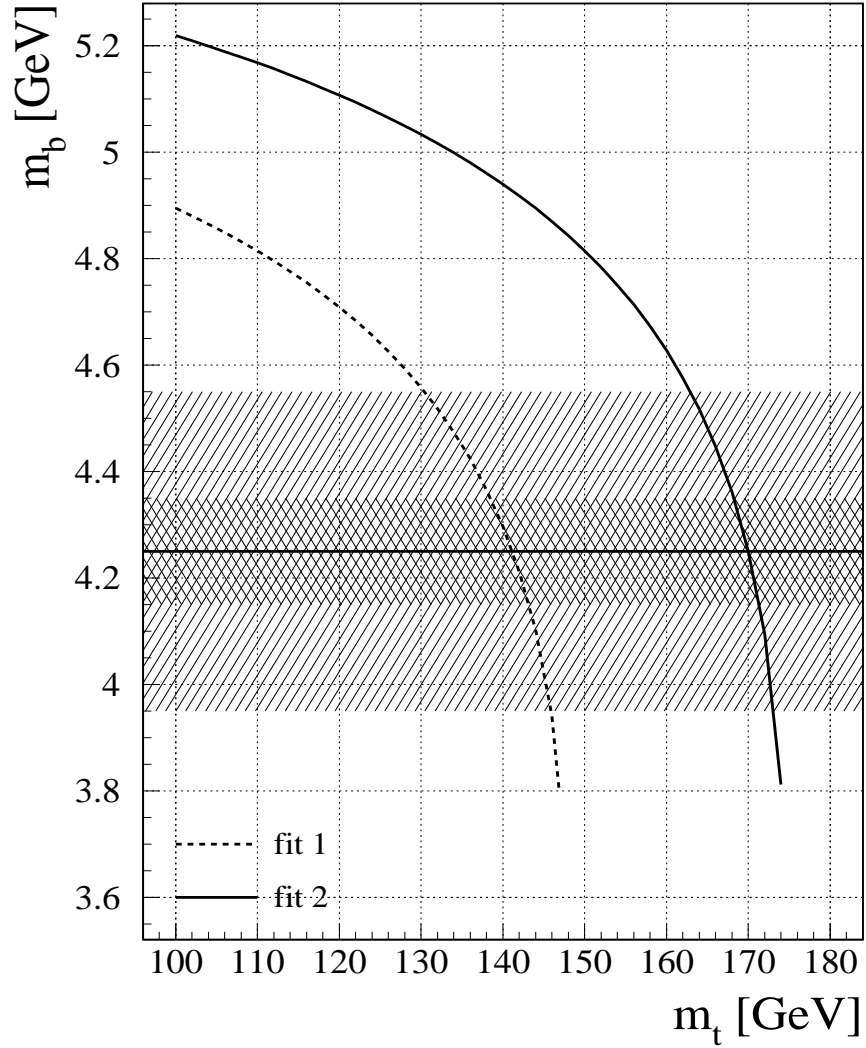}
 \end{center}
 \caption{\label{f3}
  The correlation between $\mb$ and $\mt$ for $\mze=500$ GeV.
  The hatched area indicates the experimental value for $\mb=4.25
  \pm 0.3$ GeV (the crossed area for $\mb=4.25 \pm 0.1$ GeV)
  and the curved lines give the correlation for different
  values of the coupling constant at the GUT scale.}
\end{figure}
\clearpage

\begin{figure}
 \begin{center}
  \leavevmode
  \epsfxsize=14cm
  \epsfysize=18cm
  \epsffile{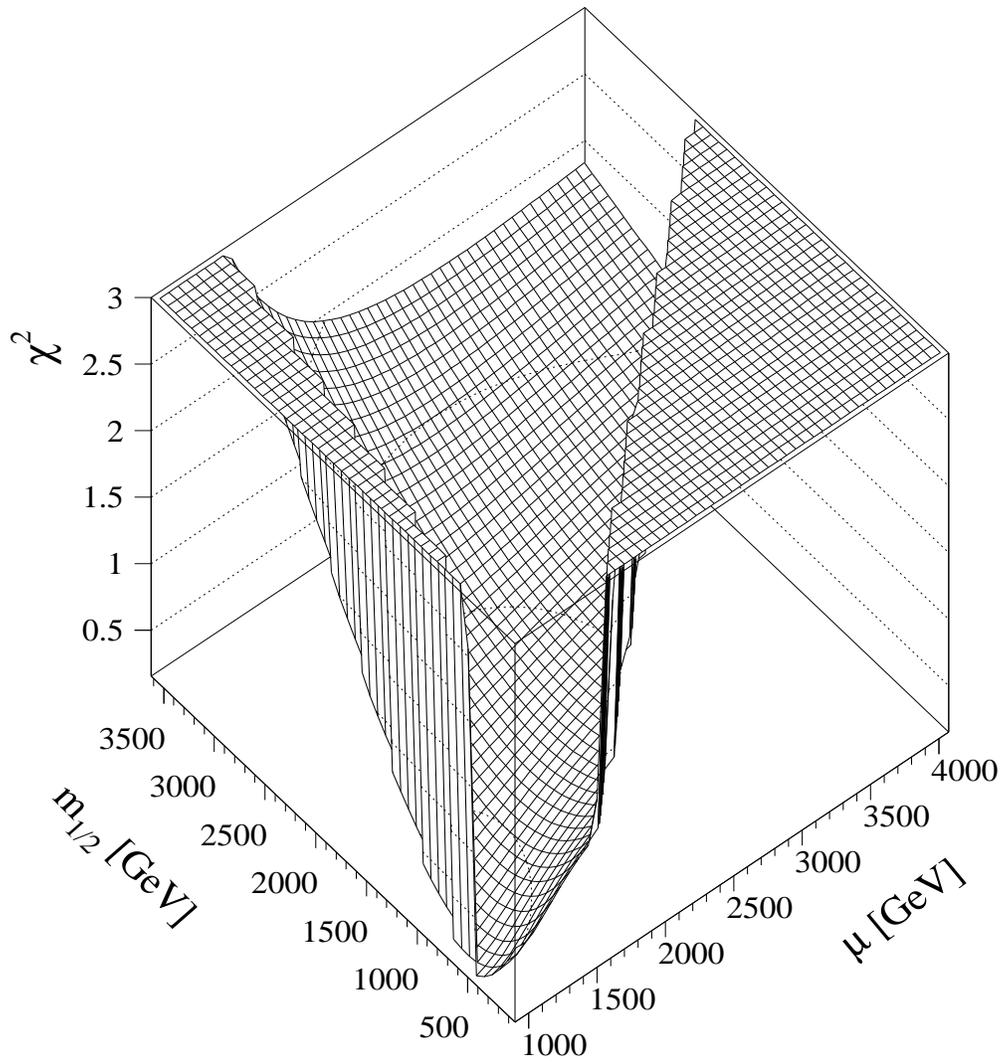}
 \end{center}
  \caption{\label{f4}
   The correlation between
   $\mha$ and $\mu$ for $\mze$=500 GeV. The vertical axis
   gives the value of $\chi^2$.  }
\end{figure}
\clearpage

\begin{figure}
 \begin{center}
  \leavevmode
  \epsfxsize=15cm
  \epsfysize=18cm
  \epsffile{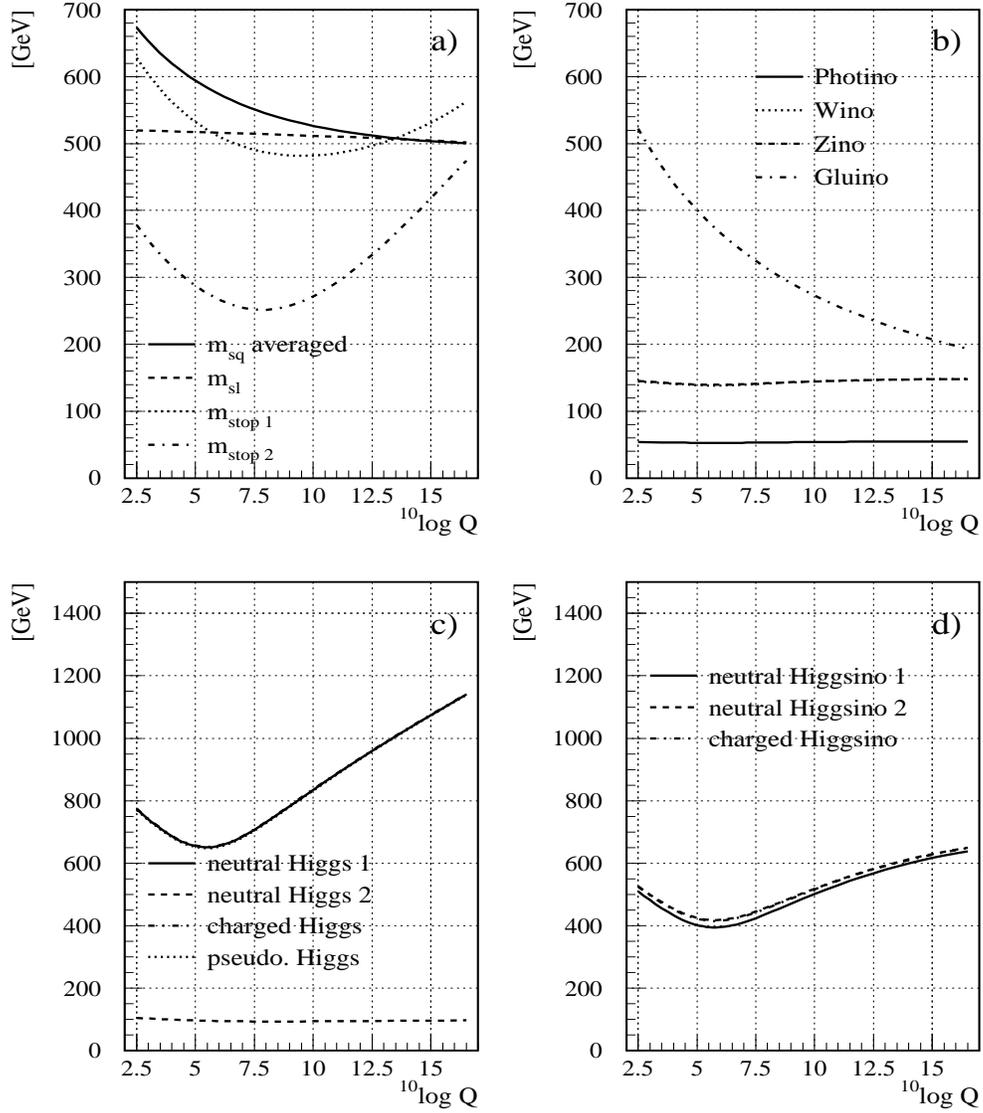}
 \end{center}
 \caption{ \label{f5}
  The evolution of masses  of the SUSY particles.
  The lines indicated as photino and zino are the lines corresponding
  to the two lightest neutralinos, while the lines of wino and
   charged higgsino correspond to the two lightest charginos.
             The lightest Higgs (neutral Higgs 2 in lower left corner)
   gets its small mass only after spontaneous symmetry breaking via
   the Higgs mechanism.}
\end{figure}
\clearpage

\begin{figure}
 \begin{center}
  \leavevmode
  \epsfxsize=14cm
  \epsfysize=18cm
  \epsffile{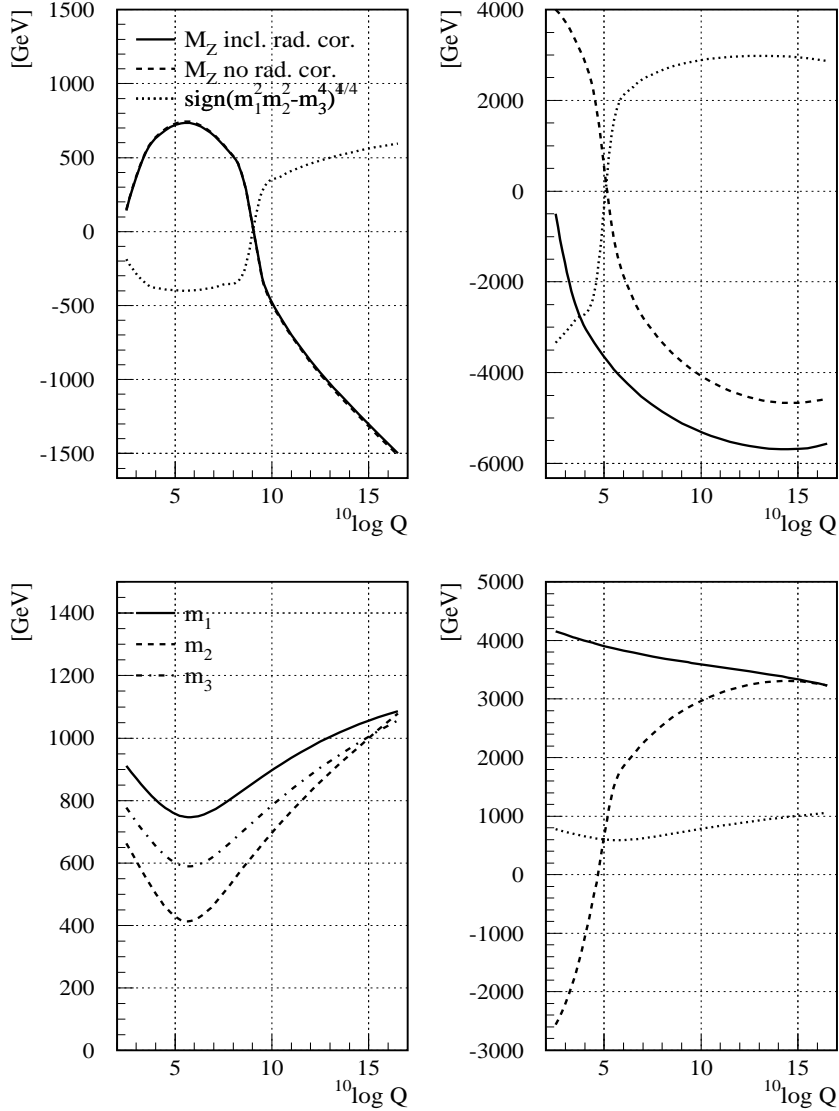}
 \end{center}
 \caption{\label{f7}
   The evolution of the \mz~mass, the determinant in the Higgs
   potential, and the value of the Higgs mass squared
   for the optimum solution (left) and the solution
   corresponding to the 90\% C.L. upper limits (right).
   The dashed lines in the top curves show the solution of the
   minimization of the Higgs potential at the tree level.
   This line coincides with the solid line for the best solution,
   as expected if supersymmetry is not badly broken, but
             the  one-loop corrections to the \mz~  masses
   become unnaturally large for the right hand side. }
\end{figure}
\clearpage
\begin{figure}
 \begin{center}
  \leavevmode
  \epsfxsize=14cm
  \epsfysize=14cm
  \epsffile{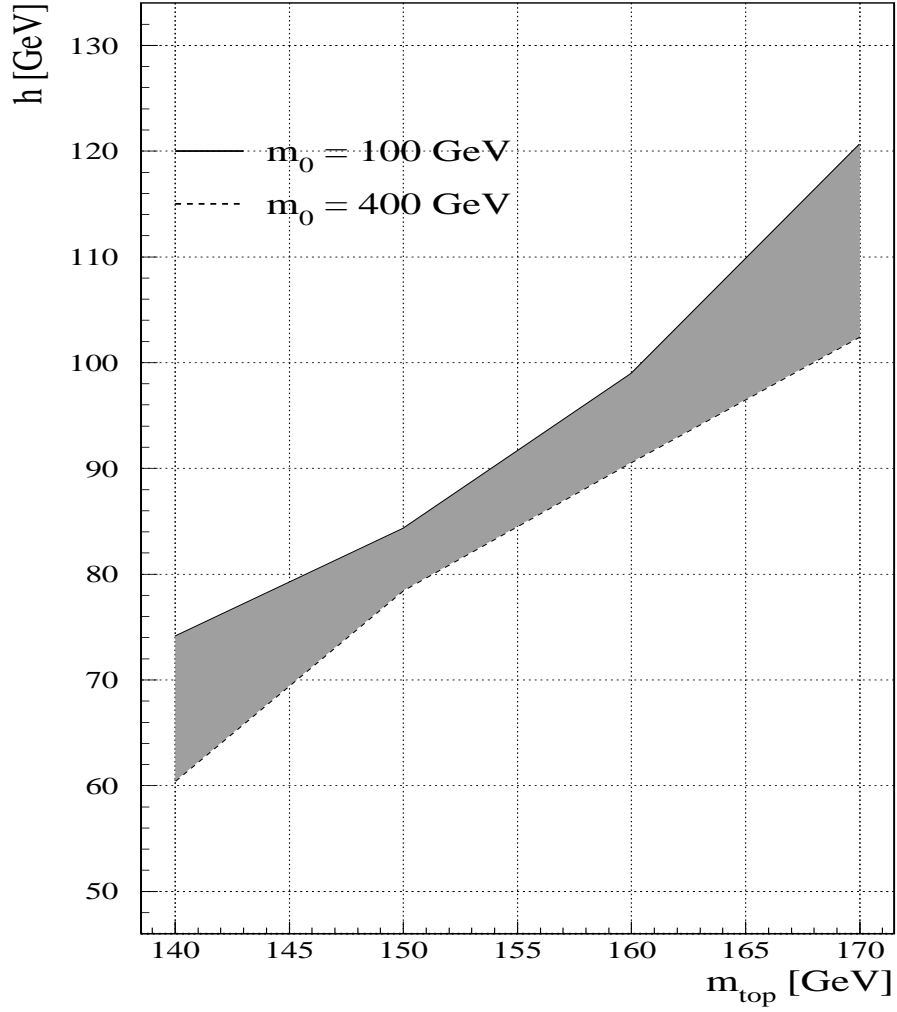}
 \end{center}
 \caption{\label{f6}
  The mass of the lightest Higgs particle as function of the top quark
  mass. The parameters  of \tb, \mze, and \mha~were optimised
  for each choice of \mt~and \mze.}
\end{figure}
\end{document}